\documentclass[lettersize,journal]{IEEEtran}
\usepackage{amsmath,amsfonts}
\usepackage{array}
\usepackage{textcomp}
\usepackage{stfloats}
\usepackage{url}
\usepackage{verbatim}
\usepackage{graphicx}
\usepackage{cite}
\usepackage{graphicx}
\usepackage{booktabs}
\usepackage[misc]{ifsym}
\usepackage{lipsum}  
\usepackage[accsupp]{axessibility}
\usepackage{listings}
\usepackage[ruled,linesnumbered]{algorithm2e}
\SetKwComment{Comment}{/* }{ */}
\usepackage{amsmath}  % add by shengxuhan
\usepackage{multicol}
\usepackage{multirow}
\usepackage{makecell}
\usepackage{adjustbox}
\usepackage{bbding}
\usepackage{orcidlink}
\usepackage{amsfonts}
\usepackage{nicefrac}
\usepackage{microtype}
\usepackage[dvipsnames]{xcolor}
\usepackage {pifont}
\usepackage{subcaption}
\usepackage{colortbl}
\definecolor{lightgray}{gray}{0.95}
\definecolor{color3}{gray}{0.95}
\definecolor{rouse}{rgb}{0.981,0.961,0.941}
\usepackage{float}
\usepackage{wrapfig}
\usepackage{subcaption}

\newcommand{\boldblue}[1]{\textcolor{Blue}{\textbf{#1}}}
\newcommand{\boldred}[1]{\textcolor{Red}{\textbf{#1}}}
\newcolumntype{T}{@{\hspace{0em}}c@{\hspace{0em}}}

\begin{document}

\title{RealOSR: Latent Guidance Boosts Diffusion-based Real-world Omnidirectional Image Super-Resolutions}

\author{
Xuhan~Sheng, Runyi~Li, Bin~Chen, Weiqi~Li, Xu~Jiang, and Jian~Zhang,~\IEEEmembership{Member,~IEEE}
\thanks{This work was supported by National Natural Science Foundation of China under Grant 62372016.}
\thanks{Xuhan Sheng, Runyi Li, Bin Chen, Weiqi Li, Xu Jiang, and Jian Zhang are with the School of Electronic and Computer Engineering, Peking University, Shenzhen 518055, China. (e-mail: shengxuhan@stu.pku.edu.cn; lirunyi@stu.pku.edu.cn; chenbin@stu.pku.edu.cn; liweiqi@stu.pku.edu.cn; xjiang25@stu.pku.edu.cn
; zhangjian.sz@pku.edu.cn). (\textit{Corresponding author: Jian~Zhang.})}
}

\markboth{Journal of \LaTeX\ Class Files,~Vol.~14, No.~8, August~2021}%
{Shell \MakeLowercase{\textit{et al.}}: A Sample Article Using IEEEtran.cls for IEEE Journals}

\IEEEpubid{\begin{minipage}{\textwidth}\ \centering
		Copyright \copyright 2025 IEEE. Personal use of this material is permitted. \\
		However, permission to use this material for any other purposes must be obtained 
		from the IEEE by sending an email to pubs-permissions@ieee.org.
\end{minipage}}
\IEEEpubidadjcol


\maketitle

\begin{abstract}
Omnidirectional image super-resolution (ODISR) aims to upscale low-resolution (LR) omnidirectional images (ODIs) to high-resolution (HR), catering to the growing demand for detailed visual content across a $ 180^{\circ}\times360^{\circ}$ viewport. Existing ODISR methods are limited by simplified degradation assumptions (e.g., bicubic downsampling), failing to model and exploit the real-world degradation information. Recent latent-based diffusion approaches using condition guidance suffer from slow inference due to their hundreds of updating steps and frequent use of VAE. To tackle these challenges, we propose \textbf{RealOSR}, a diffusion-based framework tailored for real-world ODISR, featuring efficient latent-based condition guidance within a one-step denoising paradigm. Central to efficient latent-based condition guidance is the proposed \textbf{Latent Gradient Alignment Routing (LaGAR)}, a lightweight module that enables effective pixel-latent space interactions and simulates gradient descent directly in the latent space, thereby leveraging the semantic richness and multi-scale features captured by the denoising UNet. Compared to the recent diffusion-based ODISR method, OmniSSR, RealOSR achieves significant improvements in visual quality and over \textbf{200$\times$} inference acceleration. Our code and models will be released upon acceptance.
\end{abstract}


\section{Introduction}
\label{sec:intro}

Omnidirectional images (ODIs) can capture a viewport range of $180^{\circ}\times360^{\circ}$, providing comprehensive visual information that enhances various applications. They have many projection methods like sphere projection (SP), equirectangular projection (ERP), cubemap projection (CMP), and tangent projection (TP). In practice, ODIs require extremely high resolution (e.g., 4K$\times$8K~\cite{ai2022deep}) to display details in a narrow field of view (FoV). Therefore, omnidirectional image super-resolution (ODISR) is proposed to reduce the industrial cost of high-precision camera sensors by super-resolving low-resolution (LR) ODIs into high-resolution (HR) ODIs.

Diffusion models have emerged as next-generation visual generative models. From the pioneering work~\cite{sohl2015deep}, these models are further developed by DDPM~\cite{ho2020denoising} and SDE~\cite{song2020score}. Benefiting from diffusion-based image prior, the research of planar image super-resolution (PISR) has shifted the focus from simple degradations (e.g. bicubic downsampling) to real-world degradations, which has significantly developed real-world planar image super-resolution (Real-PISR) \cite{StableSR_Wang_Yue_Zhou_Chan_Loy_2023, yang2023pasd, wu2024seesr, 2024s3diff}. Most of them use the high-order degradation pipeline of Real-ESRGAN \cite{wang2021realesrgan} to simulate real-world degradations. Current methods for Real-PISR primarily focus on planar images, with limited exploration of ODIs. Existing approaches for ODISR still face unsatisfactory performance. Most methods are end-to-end trained using pixel-level reconstruction loss (e.g. $L_1$ and $L_2$ losses) \cite{osrt_Yu_Wang_Cao_Li_Shan_Dong_2023, Deng_Wang_Xu_Guo_Song_Yang_2021}, resulting in over-smoothed and distorted outputs. In addition, they use simple ERP-bicubic or fisheye-bicubic degradation, ignoring a range of unknown degradations that real-world omnidirectional camera sensors may encounter, struggling to achieve photo-realistic SR.

Recent diffusion-based condition guidance PISR approaches~\cite{wang2022ddnm, kawar2022denoising, chung2023diffusion, fei2023generative, Rout_2023_psld, dkp_yang2024dynamic} leverage LR images as condition guidance during the sampling process. For the omnidirectional setting, OmniSSR~\cite{li2024omnissr} adapts planar-image generative priors for ODISR without training. While these methods demonstrate strong performance, they exhibit several critical limitations, as summarized in Tab.~\ref{tab:idea_comparison}. Among these, the following issues warrant particular attention.
\textit{\textbf{First}}, they assume a linear and fully known degradation operator $\mathbf{A}$, which does not hold for real-world degradations that are typically nonlinear and unknown.  
\textit{\textbf{Second}}, condition guidance must be computed in the pixel space, necessitating repeated and computationally expensive latent-pixel space conversions via a VAE.  
\textit{\textbf{Third}}, the guidance on pixel space requires ERP$\leftrightarrow$TP transformation, which is time-consuming.

\IEEEpubidadjcol

Delving into the real-world ODISR (Real-ODISR) task, from experimental observations we draw our motivation as follows: (1) An approximate degradation guidance still leads to a relatively significant performance improvement. This indicates that degradation information used for condition guidance does not need to match the input LR image strictly and perfectly. Roughly accurate degradation information can still effectively guide the generation process. (2) In the SR task with bicubic degradation, $\mathbf{A}$ and its inverse operator $\mathbf{A^{\dagger}}$ used for condition guidance \cite{wang2022ddnm,chung2023diffusion,li2024omnissr} can be approximately regarded as a downsampler and upsampler, respectively. Such up/down-sampling in the latent space still preserves most of the original image information, similar to these operations in pixel space, suggesting that operations of $\mathbf{A}$, $\mathbf{A^{\dagger}}$ could potentially be applied in the latent space, bypassing the need for pixel space.

\begin{table*}[t]
% \vspace{-2mm}
\caption{Comparison of diffusion models in conditional gradient guidance. “$\checkmark$” indicates feature support. ``×'' indicates lack of support. ``–'' denotes cases where VAE is not used for guidance. ``Backprop'' denotes back propagation. Conditional gradient guidance refers to leveraging a mathematically defined degradation model (e.g., $\mathbf{y}=\mathbf{Ax}$) to compute gradients, guiding the generation toward consistency with the observed input.}

\resizebox{\linewidth}{!}{
\begin{tabular}{l | c c c c c c c c}
\toprule
Supported Feature & DDRM~\cite{kawar2022denoising} & Diff-DKP~\cite{dkp_yang2024dynamic} & DPS~\cite{chung2023diffusion} & GDP~\cite{fei2023generative} & PSLD~\cite{Rout_2023_psld} & SeeSR~\cite{wu2024seesr} & OSEDiff~\cite{wu2024osediff} & RealOSR(Ours) \\
% Supported Feature & DDNM\cite{wang2022ddnm} & Diff-DKP\cite{dkp_yang2024dynamic} & DPS\cite{chung2023diffusion} & GDP\cite{fei2023generative} & PSLD\cite{Rout_2023_psld} & OmniSSR\cite{li2024omnissr}  & RealOSR \\
\midrule

Latent Denoising  & × & × & × & × & $\checkmark$ & $\checkmark$ & $\checkmark$   & \boldred{$\checkmark$} \\
One-Step Denoising  & × & × & × & × & × & × & $\checkmark$   & \boldred{$\checkmark$} \\
% Guidance for linear degradation  & $\checkmark$ & $\checkmark$ & $\checkmark$ & $\checkmark$ & $\checkmark$  & $\checkmark$ \\
Nonlinear Degradation Guide  & × & × & $\checkmark$ & $\checkmark$ & ×  & × & ×  & \boldred{$\checkmark$} \\
Unknown Degradation Guide & × & $\checkmark$ & × & $\checkmark$ & × & ×  & × & \boldred{$\checkmark$} \\
% Unknown Degradation Estimation & × & $\checkmark$ & × & $\checkmark$ & × & ×  & \boldred{$\checkmark$} \\
% Pixel Guidance  & $\checkmark$ & $\checkmark$ & $\checkmark$ & × & ×  & $\checkmark$ \\
Latent Guidance  & × & × & × & × & $\checkmark$ & × & ×   & \boldred{$\checkmark$} \\
Guide w/o VAE Backprop & - & - & - & - & × & - & -   & \boldred{$\checkmark$} \\
Guide w/o VAE & - & - & - & - & × & - & - & \boldred{$\checkmark$} \\

\bottomrule
\end{tabular}
}
\label{tab:idea_comparison}
% \vspace{-5pt}
% \vspace{-5mm}
\end{table*}

\begin{figure*}[h]  % htbp
    % \vspace{-4mm}
    \centering
    \begin{subfigure}[b]{0.56\textwidth}
        \centering
        \includegraphics[width=\textwidth]{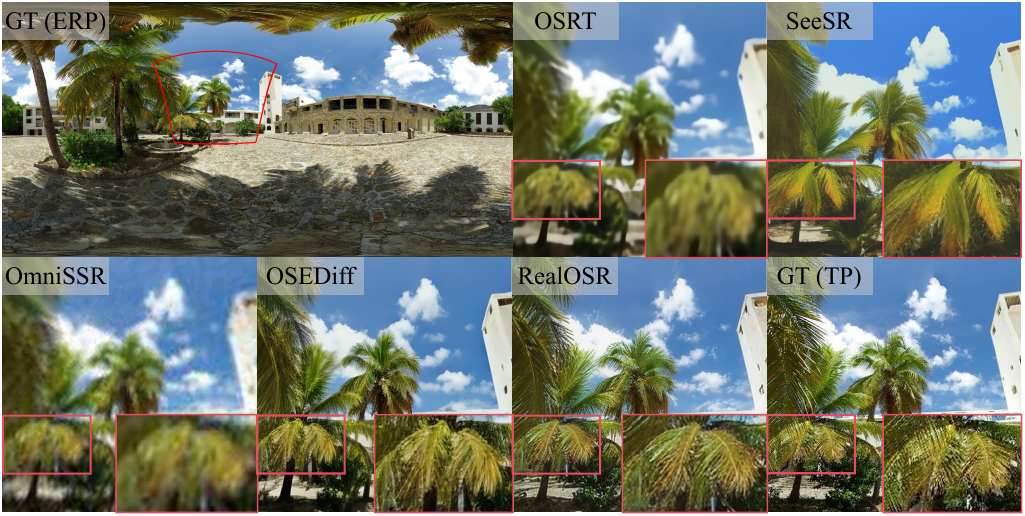}
        \caption{Visualization Comparison}
        % \caption{ERP and TP format of ODI-SR test-set image (id=20), and its restored results under real-world degradation. Zoom in for more detail.}
        \label{fig:subfig1}
    \end{subfigure}
        \hfill
        % \hspace{1cm}
    \begin{subfigure}[b]{0.42\textwidth}
        \centering
        \includegraphics[width=\textwidth]{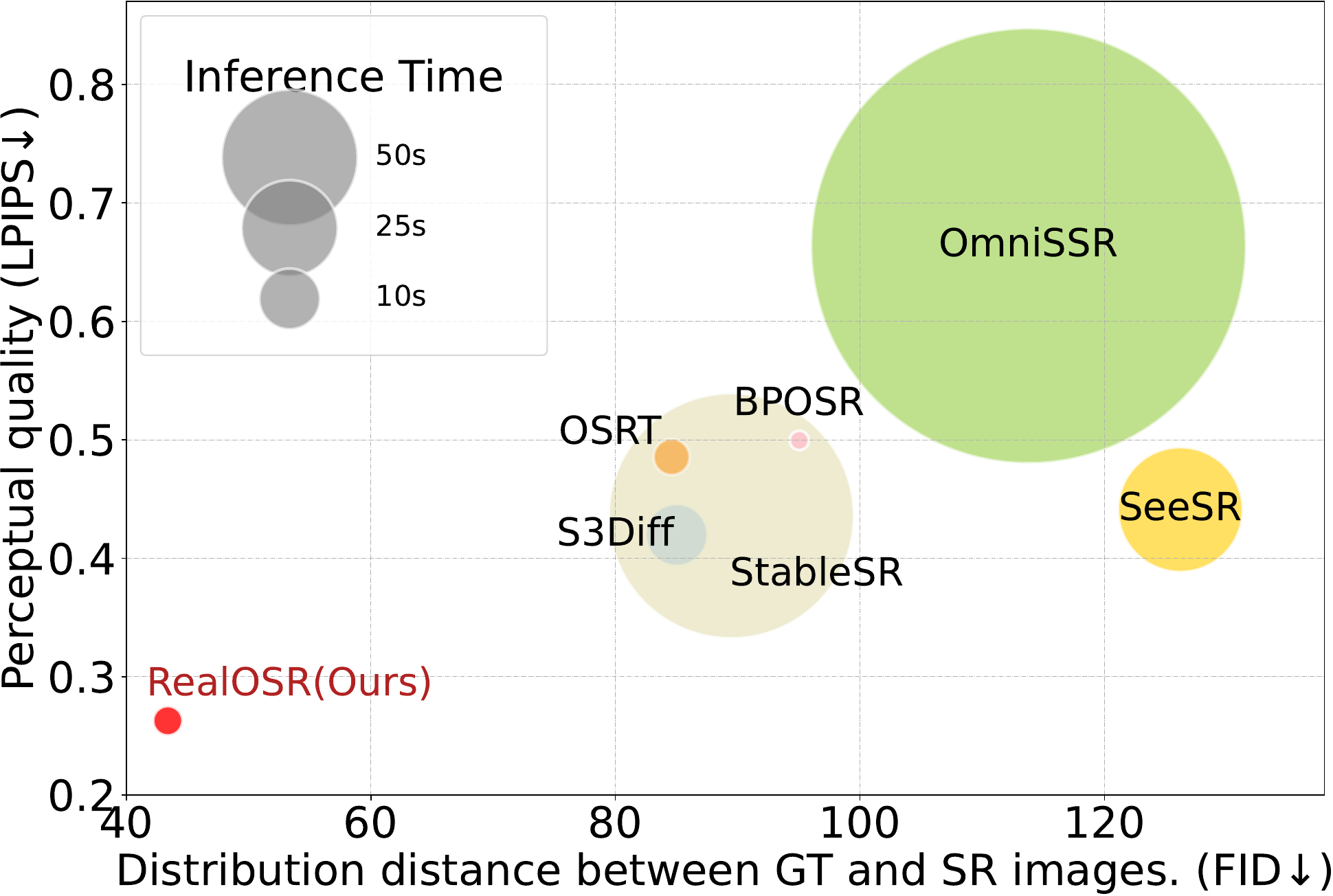}
        \caption{Comparison of image quality and efficiency}
        % \caption{Metrics among current SR approaches and our method. We standardize the data as some metrics like FID lead to better results with a lower value.}
        \label{fig:subfig2}
    \end{subfigure}
    % \vspace{-1mm}
\caption{Comparison of RealOSR and other approaches demonstrates superior performance of both fidelity and visual realness of RealOSR. (a) ERP and TP format of ODI-SR test-set image (id=20), and its restored results under real-world degradation. Zoom in for more detail. (b) RealOSR achieves comparable speed to end-to-end methods (OSRT~\cite{osrt_Yu_Wang_Cao_Li_Shan_Dong_2023}, BPOSR~\cite{wang2024bposr}) and is over \textbf{200$\times$} faster than the diffusion-based OmniSSR~\cite{li2024omnissr}, while producing substantially better visual quality.}
    \label{fig:teaser}
    \vspace{-4mm}
\end{figure*}

Based on the aforementioned priors and insights, we propose \textbf{RealOSR}, a one-step denoising diffusion model designed to handle Real-ODISR task. Specifically, our method introduces the following key innovations:
(1) \textbf{Integration of Real-World Degradation}: We incorporate real-world degradation into the ODISR task by developing degradation-aligned LR-HR image pairs for training and evaluation, ensuring that our model is trained and tested under conditions closely matching real-world scenarios. (2) \textbf{Latent Gradient Alignment Routing}: We propose a novel latent-space guidance module that simulates degradation-aware optimization dynamics by routing gradient-aligned information through the hierarchical feature spaces of the denoising UNet. Instead of relying on explicit pixel-space, we leverage rich semantic and multi-scale features in the denoising UNet, effectively improving the fidelity and perceptual quality of the reconstructed images.
(3) \textbf{Efficient Condition Guidance within One-Step Sampling}: RealOSR employs an efficient diffusion-based SR model that integrates condition guidance in one-step sampling, significantly reducing inference time. This efficiency accelerates the processing and makes our method suitable for real-world applications where timely results are crucial (in Fig.~\ref{fig:teaser}). To summarize, our contributions are listed as follows:

\noindent \ding{113}~(1) Toward real-world omnidirectional image super-resolution, we propose \textbf{RealOSR}, a diffusion model with efficient condition guidance through a one-step denoising process, and construct Real-ODISR datasets for effective model training and evaluation.

\noindent \ding{113}~(2) We introduce \textbf{Latent Gradient Alignment Routing (LaGAR)}, a lightweight module designed to guide the denoising process by aligning latent-space gradients with degradation priors.

\noindent \ding{113}~(3) LaGAR incorporates two key modules: \textbf{Latent-Pixel Transcoding Bridge} that enables efficient pixel-latent feature mapping and \textbf{Latent Gradient Simulation Core} that simulates gradient descent directly in latent space.

\noindent \ding{113}~(4) Extensive experiments on constructed benchmarks demonstrate that RealOSR achieves superior visual quality and over \textbf{200$\times$} faster inference compared to recent diffusion-based methods, validating its effectiveness and efficiency.

\section{Related Work}
\label{sec:formatting}
% \subsection{Omnidirectional Image Super-Resolution}
\textbf{Planar Image Super-Resolution (PISR).}
PISR methods fall into two categories: regressive and generative. The former~\cite{mou2022metric, Chen_2023_ICCV, Zhang_2021_ICCV, Lugmayr_2020_CVPR_Workshops, Lu_2022_CVPR, li2022d3c2, zhang2022herosnet, cheng2023hybrid, hu2019channel, guan2024frequency, zhang2021two} treats PISR as pixel-wise regression optimized with L1/L2 loss, often leading to over-smoothed results. The latter, leveraging perceptual-oriented loss~\cite{liu2020photo, niu2024learning}, GANs~\cite{han2025unsupervisedface, wang2021realesrgan, chen2023dynamic} and diffusion models~\cite{wang2022ddnm, chung2023diffusion, chung2022improving, Song_Zhang_Yin_Mardani_Liu_Kautz_Chen_Vahdat, fei2023generative, sr3, StableSR_Wang_Yue_Zhou_Chan_Loy_2023, yang2023pasd, wu2024seesr, yu2024scaling, sun2023ccsr, yang2024motion, korkmaz2025leveraging}, learns HR image distributions for more visually appealing results. Meanwhile, the research focus has shifted from simple degradations (e.g., bicubic) to real-world ones, with Real-ESRGAN~\cite{wang2021realesrgan} introducing a widely adopted degradation pipeline to simulate real-world degradations. However, GAN-based methods suffer from training instability and visual artifacts. 
Diffusion-based methods require multi-step denoising for stable, high-quality generation but are significantly slower than single-forward approaches.
PASD \cite{yang2023pasd} integrates low- and high-level features via pixel-aware cross-attention. 
SUPIR \cite{yu2024scaling} leverages the powerful priors from LLaVA \cite{liu2024visual} and SDXL~\cite{Podell_English_Lacey_Blattmann_Dockhorn_Müller_Penna_Rombach}. 
SeeSR \cite{wu2024seesr} employs tag-style text prompts to guide the diffusion process. As a compromise between GAN-based and diffusion-based approaches, one-step denoising PISR methods leverage diffusion priors with GAN-like training losses. OSEDiff~\cite{wu2024osediff} employs LoRA~\cite{hu2022lora} to adapt Stable Diffusion for one-step denoising PISR, while S3Diff~\cite{2024s3diff} further introduces degradation-aware LoRA for better adaptation to varying degradation levels.

\noindent \textbf{Omnidirectional image super-resolution (ODISR).} 
As an extension of PISR, ODISR aims to super-resolve ODIs by addressing its unique challenges~\cite{arican2011joint,sun2023opdn,9506233,10222760, ozcinar2019super, Fakour-Sevom_Guldogan_Kamarainen_2018, liu2025diffosr, yang2025geometric, wen2025mambaosr, ai2024dream360, wang2024egocentric}. ODISR methods can also be categorized into regressive and generative. 
% Kämäräinen et al.~\cite{Fakour-Sevom_Guldogan_Kamarainen_2018} leverage convolutional neural networks to effectively upscale LR omnidirectional images while preserving spatial details. Smolic et al.~\cite{ozcinar2019super} introduce GANs to enhance the visual quality of ODIs by hallucinating high-frequency details.
LAU-Net~\cite{Deng_Wang_Xu_Guo_Song_Yang_2021} is a CNN structure ODISR method with adaptive latitude band selection trained by reinforcement learning.
SphereSR~\cite{Yoon_Chung_Wang_Yoon} addresses non-uniformity in different projections by learning upsampling processes and ensuring information consistency using LIIF~\cite{Chen_Liu_Wang_2021}. Considering the difference in data representation between ODIs and planar images, a spherical pseudo-cylindrical representation\cite{cai2024spherical} is proposed, model-agnostic to most off-the-shelf SR methods, and enhances their performances. OSRT~\cite{osrt_Yu_Wang_Cao_Li_Shan_Dong_2023} and OPDN~\cite{sun2023opdn} design a distortion-aware transformer to modulate ERP distortions. BPOSR~\cite{wang2024bposr} is a bi-projection transformer considering ERP horizontal similarity and CMP perspective variability. Research on diffusion-based ODISR is limited. OmniSSR~\cite{li2024omnissr} applies projection transformation to harness pre-trained diffusion-based PISR methods, incorporating Gradient Decomposition as a condition guidance for ODISR. Despite promising advancements, real-world ODISR settings remain largely unexplored.

\section{Method}
\label{sec:method}

\subsection{Preliminaries}
\label{sec:preliminaries}
\textbf{Diffusion-based Image Inverse Problem Solving.}
Image linear degradation can be modeled as $\mathbf{y}=\mathbf{Ax}+\mathbf{n}$, where $\mathbf{y}$ denotes the degraded image, $\mathbf{A}$ the degradation operator, $\mathbf{x}$ the original image, and $\mathbf{n}$ the random noise. 
The \textit{image inverse problem} is to solve $\mathbf{x}$, formulated as:
% and we can solve it using convex optimization techniques. Specifically, our optimization objective is formulated as:
\begin{equation}
\label{eq:argmin}
    \underset{\mathbf x}{\arg\min}||\mathbf{y}-\mathbf{Ax}||_{2}^{2}+\lambda \mathcal{R}(\mathbf{x}),
\end{equation}
where $||\mathbf{y}-\mathbf{Ax}||_2^2$ ensures that the consistency with the degraded image, and the regularization term $\mathcal{R}(\mathbf{x})$ enforces image priors on $\mathbf{x}$, such as smoothness and sparsity. Diffusion models can act as a powerful regularization term, complemented by condition guidance for controllable generation~\cite{wang2022ddnm, chung2023diffusion}. 
Starting from random noise $\mathbf{x}_{{T}}$, they iteratively denoise $T$ steps to generate a clean image $\mathbf{x}_{0}$ with a denoiser $\boldsymbol{\epsilon}_{\theta}$.
At time step $t$, $\mathbf{x}_{0}$ is estimated as $\mathbf{x}_{0|t}$ from noised image $\mathbf{x}_{t}$, and further corrected with $\mathbf{y}$ and $\mathbf{A}$ as follows:
\begin{equation}
\label{eq:x_0|t}
     \mathbf{x}_{0|t} = \frac{1}{\sqrt{\overline{\alpha}_t}}(\mathbf{x}_t + \big(1-\overline{\alpha}_t)\boldsymbol{\epsilon}_{\theta}(\mathbf{x}_t, t) \big),
\end{equation}
\begin{equation}
\label{eq:dps}
     \hat{\mathbf{x}}_{0|t} = \mathbf{x}_{0|t} - \alpha \nabla_{\mathbf{x}_{0|t}} ||\mathbf{y}-\mathbf{A}\mathbf{x}_{0|t}||_2^2,
\end{equation}
where $\hat{\mathbf{x}}_{0|t}$ denotes the corrected estimate, and $\alpha$ the step size for gradient descent. For linear $\mathbf{A}$, Eq.~\ref{eq:dps} simplifies to:
\begin{equation}
    \hat{\mathbf{x}}_{0|t} = \mathbf{x}_{0|t} - 2\alpha(\mathbf{A}^{\dagger}\mathbf{A}\mathbf{x}_{0|t} - \mathbf{A}^{\dagger}\mathbf{y}).
\label{eq:gd}
\end{equation}
Then, $\mathbf{x}_{t-1}$ is sampled from $ p(\mathbf{x}_{t-1}|\mathbf{x}_{t},\hat{\mathbf{x}}_{0|t})$. Repeating the above steps yields the final result ${\mathbf{x}}_{0}$.

\textbf{Bridging Domain Gap with Projection Transformation.}
ERP images exhibit significant distortions and deformation in polar regions, which prevents them from enjoying planar image priors for ODISR. To bridge this domain gap, we can convert ERP images into TP images, which conform to the distribution of planar images~\cite{li2024omnissr}.

For a pixel $P_e(x_e, y_e)$ within the ERP image, we first find its corresponding pixel $P_s(\theta, \phi)$ on the unit sphere using Eq.~\ref{eq:erp2sphere}:
\begin{equation}
    % \lambda == \theta 
    \theta = 2\pi x_e / W,\; \phi = \pi y_e / H,
\label{eq:erp2sphere}
\end{equation}
where $H$ and $W$ are the height and width of the ERP image.
The Cartesian coordinates of the ERP image and the angular coordinates on the unit sphere exhibit a straightforward one-to-one linear relationship, suggesting a conceptual equivalence between these two projection formats.

Given the spherical coordinates of the tangent plane center $(\theta_c, \phi_c)$, The transformation from $P_s(\theta, \phi)$ to $P_t(x_t, y_t)$, i.e. ERP$\rightarrow$TP, is defined as:
\begin{equation}
\begin{aligned}
    &x_t = \big( \cos(\phi)\sin(\theta - \theta_c) \big) \big/ \zeta , \\
    &y_t = \big( \cos(\phi_c)\sin(\phi) - \sin(\phi_c)\cos(\phi)\cos(\theta - \theta_c) \big) \big/ \zeta , \\
    &\zeta = \sin(\phi_c)\sin(\phi) + \cos(\phi_c) \cos(\phi) \cos(\theta - \theta_c).
\end{aligned}
\label{eq:sphere2tan}
\end{equation}
% where $\zeta = \sin(\phi_c)\sin(\phi) + \cos(\phi_c) \cos(\phi) \cos(\theta - \theta_c)$.
The corresponding inverse transformation, i.e. TP$\rightarrow$ERP, is:
\begin{equation}
\begin{aligned}
    &\theta = \theta_c + \arctan \big( \frac{x_t\sin(c)}{\rho\cos(\phi_1)\cos(c) - y_t\sin(\phi_c)\sin(c)}\big), \\
    % &\theta = \theta_c + \arctan (\frac{x_t\sin(c)}{\rho\;\cos(\phi_1)\cos(c) - y_t\;\sin(\phi_c)\sin(c)}) \\
    &\phi = \arcsin \big( \cos(c)\sin(\phi_c) +  y_t  \sin(c)\cos(\phi_c) / \rho \big),
\end{aligned}
\label{eq:tan2sphere}
\end{equation}
where $\rho = \sqrt{x_t^2 + y_t^2}$ and $c=\arctan(\rho)$.

With Eq.~\ref{eq:sphere2tan} and Eq.~\ref{eq:tan2sphere}, we can build one-to-one forward and inverse mapping functions between pixels on the ERP image and pixels on the TP images.

\subsection{Motivation}
\label{sec:motivation}
This section introduces our motivation for supporting conditional gradient guidance in latent space:

\noindent \ding{113}~\textbf{Condition Guidance in Latent Diffusion.} Previous latent-based diffusion models, such as PSLD \cite{Rout_2023_psld}, perform denoising in the latent space and use VAE to convert the results to pixel space for gradient updates, requiring VAE backpropagation and incurring additional memory and time costs. OmniSSR \cite{li2024omnissr} introduces a condition-guidance technique that avoids backpropagation, but still uses VAE for latent and pixel space conversion at each step.

\begin{figure}[h]
    \centering
    \includegraphics[width=\linewidth]{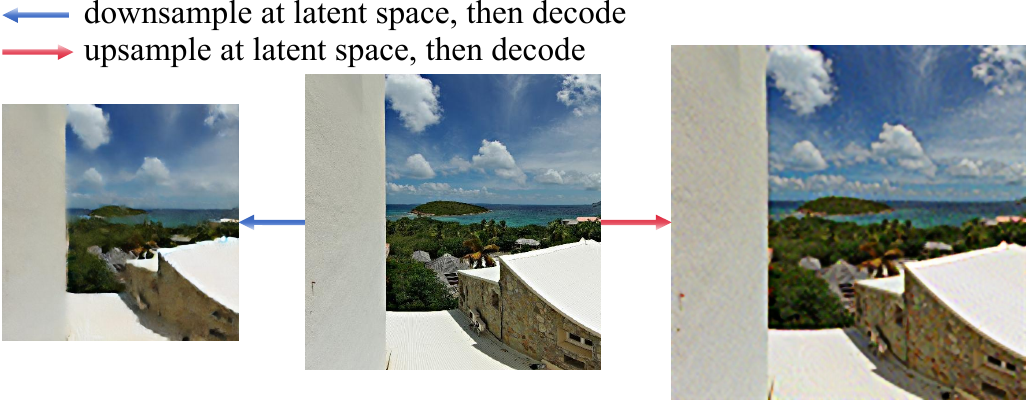}
    % \vspace{-5mm}
    \caption{Bicubic up/down-sampling in latent space, decoded to pixel space for visualization. Despite minor damage, the original information is largely preserved. This motivates direct condition guidance in latent space.}
    \label{fig:motivation}
  \vspace{-4mm}
\end{figure}

\noindent \ding{113}~\textbf{Degradation Correspondence in Latent Space.} Latent and pixel spaces, though distinct, exhibit strong correspondence. In SR tasks with bicubic downsampling, the degradation operator $\mathbf{A}$ and its pseudo-inverse $\mathbf{A}^{\dagger}$ correspond to specific downsampling and upsampling operations. We apply them directly in the latent space with results visualized in pixel space (Fig.~\ref{fig:motivation}). While the up/down-sampled images show certain damage, the semantic and structural information remains largely preserved. This motivates conditional gradient guidance in latent space, where feature representations are semantically richer, making it better suited for addressing complex, real-world, non-linear degradation.

\noindent \ding{113}~\textbf{Unknown Degradation.} A preliminary experiment was conducted (see Tab.~\ref{tab:unmatched-rnd}) with fisheye-bicubic downsampling \cite{osrt_Yu_Wang_Cao_Li_Shan_Dong_2023} to generate LR ODIs. While the degradation guidance in OmniSSR is modeled as ERP-bicubic downsampling, OmniSSR still outperforms its base model, StableSR, across WS-PSNR, WS-SSIM, and FID. This suggests that the degradation operator $\mathbf{A}$ in OmniSSR need not precisely match the input image degradation; an approximate degradation operator can still effectively guide the diffusion model to achieve better results. Inspired by this observation, we propose using a learnable module to dynamically estimate image-specific degradation conditions, supporting gradient simulation in latent space with improved speed and reduced memory.
% \vspace{-7mm}

\begin{table}[h]
    \caption{Despite fisheye-bicubic $\times$4 downsampling input, OmniSSR~\cite{li2024omnissr}, using ERP-bicubic as guidance, surpasses its base model StableSR~\cite{pan2021exploiting} in overall performance. This suggests that the degradation used for guidance does not need to precisely match the realistic degradation of the input image. Best results are shown in \textbf{bold}.}
    \centering
    \resizebox{\linewidth}{!}{
        \begin{tabular}{c|ccc}
        \toprule
        ODI-SR Fisheye-bicubic$\times$4    & WS-PSNR↑                              & WS-SSIM↑                               & FID↓                                  \\ \hline
        StableSR                           & 23.19                                 & 0.6548                                 & 56.35                                 \\ 
        OmniSSR (ERP-bicubic) &  \textbf{25.21} & \textbf{0.7136} & \textbf{37.09} \\
        \bottomrule
        \end{tabular}
    }
    \label{tab:unmatched-rnd}
    % \vspace{-4mm}
\end{table}

\begin{figure*}
    \centering
    \includegraphics[width=1\linewidth]{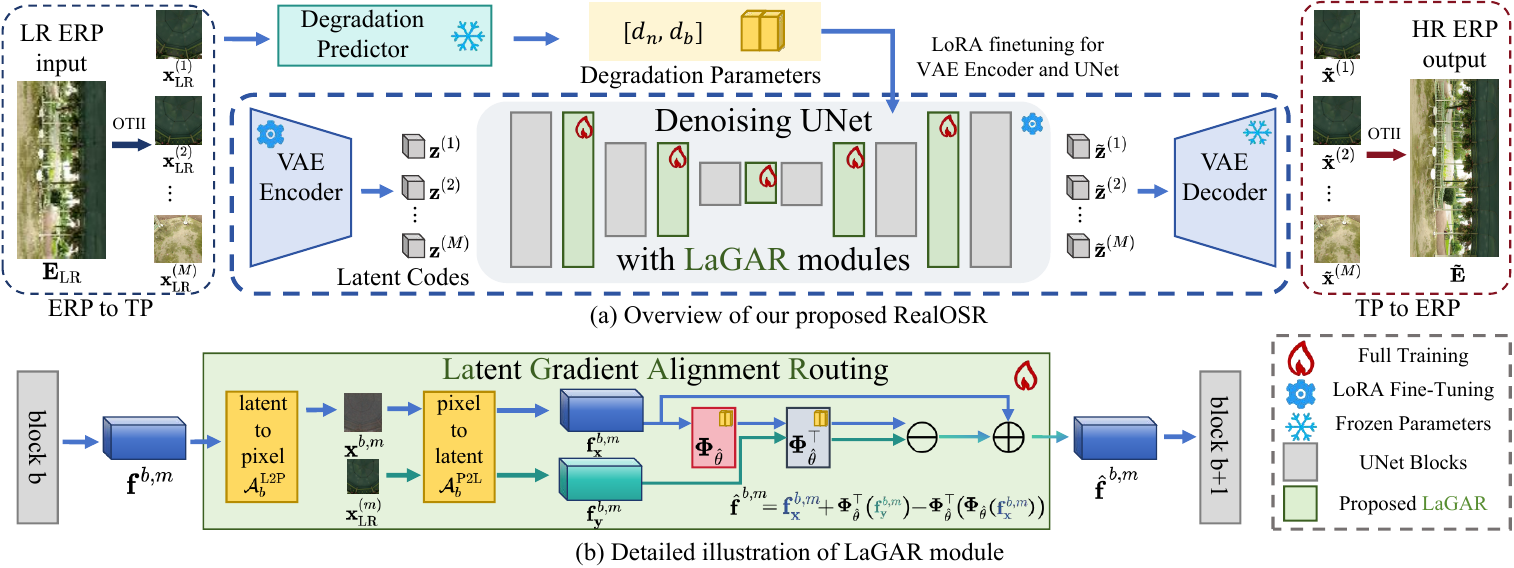}
    \caption{Overall architecture and detailed Latent Gradient Alignment Routing (LaGAR) of our proposed RealOSR. Input LR ERP is first transformed into TP images, and then sent sequentially through the SD VAE Encoder into denoising UNet with (1) degradation-aware LoRA, and (2) LaGAR for condition guidance in one-step denoising. All generated TP images are projected back to ERP format as the final SR result.}
    \label{fig:overview}
    \vspace{-4mm}
\end{figure*}

\subsection{Overview}

As illustrated in Fig.~\ref{fig:overview}, RealOSR achieves conditional gradient guidance by injecting LaGAR modules between the Stable Diffusion UNet blocks. The inference pipeline of RealOSR is presented in Algo.~\ref{alg:inference-pipeline}. First, the LR ERP image $\mathbf{E}_{\text{LR}}$ undergoes ERP$\rightarrow$TP transformation $\mathcal{F}$, generating $M$ number of TP images $\{ \mathbf{x}_\text{LR}^{(1)},\mathbf{x}_\text{LR}^{(2)},...,\mathbf{x}_\text{LR}^{(M)} \}$. Next, the degradation predictor (DP) estimates two degradation parameters, $d_n, d_b$, for each LR TP image $\mathbf{x}_\text{LR}^{(m)}$. Based on the degradation parameters, the ``LoraGenerator" generates specific LoRA weights, which are then loaded into the SD VAE Encoder $\mathcal{E}$ and SD UNet $\boldsymbol{\epsilon}_{\theta}$. The SD VAE Encoder then processes each TP image to obtain its latent code $\mathbf{z}^{(m)}$. Subsequently, UNet denoising with LaGAR produces the denoised latent code $\tilde{\mathbf{z}}^{(m)}$, which is then decoded back into the pixel space, yielding the HR TP image $\tilde{\mathbf{x}}^{(m)}$. Once all HR TP images are generated, they are transformed back into an HR ERP image $\tilde{\mathbf{E}}$ via the TP$\rightarrow$ERP transformation $\mathcal{F}^{-1}$.

The training pipeline of RealOSR largely follows the inference pipeline, as shown in Algo.~\ref{alg:training-pipeline}. Since the network computation only relates to TP images, we construct LR-HR TP image pairs for training. Thus, projection transformation is unnecessary for model training, significantly reducing memory consumption.

% \begin{figure}[H]
% \centering
% \begin{minipage}[t]{0.48\textwidth}
\begin{algorithm}
\caption{RealOSR Inference Pipeline}
\label{alg:inference-pipeline}
\KwIn{$\mathbf{E}_{\text{LR}}$, $\mathcal{F}$, $\mathcal{F}^{-1}$, DP,  $\boldsymbol{\epsilon}_{\theta}$, $\mathcal{E}$, $\mathcal{D}$, LoraGenerator, LaGAR}
\KwOut{SR result $\tilde{\mathbf{E}}$}
$\{ \mathbf{x}_\text{LR}^{(1)}, \mathbf{x}_\text{LR}^{(2)},\ldots,\mathbf{x}_\text{LR}^{(M)} \} = \mathcal{F}(\mathbf{E}_{\text{LR}})$ \\
\For{$m=1$ \KwTo $M$}{
  $d_n, d_b = \text{DP} (\mathbf{x}_\text{LR}^{(m)})$ \\
  $\theta_{\mathcal{E}}, \theta_{\text{UNet}} = \text{LoraGenerator}(d_n, d_b)$ \\
  $\mathcal{E}$.load($\theta_{\mathcal{E}}$) \\
  $\boldsymbol{\epsilon}_{\theta}$.load($\theta_{\text{UNet}}$) \\
  $\mathbf{z}^{(m)} = \mathcal{E}(\mathbf{x}_\text{LR}^{(m)})$\\
  Obtain $\tilde{\mathbf{z}}^{(m)}$ via Algo.~\ref{alg:LaGAR} with LaGAR. \\
  % Obtain $\tilde{\mathbf{z}}^{(m)}$ via UNet $\boldsymbol{\epsilon}_{\theta}$ denoising with LaGAR. \\
  $\tilde{\mathbf{x}}^{(m)} = \mathcal{D}(\tilde{\mathbf{z}}^{(m)})$
}
$\tilde{\mathbf{E}} = \mathcal{F}^{-1}(\{\tilde{\mathbf{x}}^{(1)}, \tilde{\mathbf{x}}^{(2)}, \ldots, \tilde{\mathbf{x}}^{(M)} \})$ \\
\Return $\tilde{\mathbf{E}}$
\end{algorithm}
\vspace{-6mm}
% \end{minipage}
% \begin{minipage}[t]{0.48\textwidth}
\begin{algorithm}
\caption{RealOSR Training Pipeline}
\label{alg:training-pipeline}
\KwIn{$\mathbf{x}_\text{LR}^{(m)}$, $\mathbf{x}_\text{HR}^{(m)}$, DP, $\boldsymbol{\epsilon}_{\theta}$, $\mathcal{E}$, $\mathcal{D}$, LoraGenerator, LaGAR, $\mathcal{L}_{total}$}
\KwOut{$Loss$}
$d_n, d_b = \text{DP} (\mathbf{x}_\text{LR}^{(m)})$ \\
$\theta_{\mathcal{E}}, \theta_{\text{UNet}} = \text{LoraGenerator}(d_n, d_b)$ \\
$\mathcal{E}$.load($\theta_{\mathcal{E}}$) \\
$\boldsymbol{\epsilon}_{\theta}$.load($\theta_{\text{UNet}}$) \\
$\mathbf{z}^{(m)} = \mathcal{E}(\mathbf{x}_\text{LR}^{(m)})$\\
Obtain $\tilde{\mathbf{z}}^{(m)}$ via Algo.~\ref{alg:LaGAR} with LaGAR. \\
% Obtain $\tilde{\mathbf{z}}^{(m)}$ via UNet $\boldsymbol{\epsilon}_{\theta}$ denoising with LaGAR. \\
$\tilde{\mathbf{x}}^{(m)} = \mathcal{D}(\tilde{\mathbf{z}}^{(m)})$ \\
$Loss = \mathcal{L}_{total}(\tilde{\mathbf{x}}^{(m)}, \mathbf{x}_\text{HR}^{(m)})$ \\
\textbf{Update} $\boldsymbol{\epsilon}_{\theta}$, $\mathcal{E}$, LoraGenerator, LaGAR \textbf{via backpropagation using} $Loss$ \\
\Return $\boldsymbol{\epsilon}_{\theta}$, $\mathcal{E}$, LoraGenerator, LaGAR
\end{algorithm}
% \vspace{-8mm}
% % \end{minipage}
% \end{figure}

\subsection{Latent Gradient Alignment Routing (LaGAR)}

The insights in Sec.~\ref{sec:motivation} motivate us to achieve real-world degradation-based condition guidance in the latent spaces. To further achieve one-step denoising, we refer to implementing such guidance between UNet blocks. For simplicity, here, the ``\textbf{latent spaces}'' refers to the whole feature spaces expanded by each UNet block in contrast to pixel space. The LaGAR module contains two components: (1) \textbf{Latent-Pixel Transcoding Bridg} and (2) \textbf{Latent Gradient Simulation Core}.

\noindent\textit{\textbf{Latent-Pixel Transcoding Bridge (LPTB).}} Since the pixel space of the LR image and the latent spaces expanded by the UNet blocks have a large domain gap, we propose the Latent-Pixel Transcoding Bridge composed of latent-to-pixel (L2P) and pixel-to-latent (P2L) to achieve the conversion between the latent space and the pixel space, thus supporting efficient information transformation. As the latent spaces expanded by different UNet blocks are different, a series of LPTBs are required for each UNet block. Therefore, the LPTB is designed to be parameter efficient. 

Specifically, for the $m$-th TP image $\mathbf{x}^{(m)}_{\text{LR}}$ and the 
corresponding feature map $\mathbf{f}^{b, m}$ output by the $b$-th UNet block, the conversions are achieved as follows:
\begin{equation}
\begin{aligned}
    \mathbf{x}^{b, m} = \mathcal{A}^\text{L2P}_b(\mathbf{f}^{b, m}), \\
    \mathbf{f}_\mathbf{x}^{b, m} = \mathcal{A}^\text{P2L}_b(\mathbf{x}^{b, m}), \\
    \mathbf{f}^{b, m}_\mathbf{y} = \mathcal{A}^\text{P2L}_b(\mathbf{x}^{(m)}_{\text{LR}}), 
\end{aligned}
\label{eq:dam}
\end{equation}
where $\mathcal{A}^\text{L2P}_b(\cdot)$ converts feature maps from the $b$-th UNet block into pixel space, allowing for pixel-level control if needed (e.g., gradient guidance in the pixel space). Conversely, $\mathcal{A}^\text{P2L}_b(\cdot)$ enables efficient mapping back to the latent space of the $b$-th UNet block, supporting latent-level control, such as the LGSC to be introduced in the following part. LPTB is designed to be lightweight by utilizing 1$\times$1 group convolution and Channel Shuffle~\cite{zhang2018shufflenet} for efficient conversion while employing pixel shuffle/unshuffle operations for fast up/down-sampling.

\noindent\textit{\textbf{Latent Gradient Simulation Core (LGSC).}} Degradation-based condition guidance is trivial when $\mathbf{A}$ is linear and known. In real-world ODISR, however, the degradation operator $\mathbf{\Phi}$ is non-linear and unknown, making gradient calculation in Eq.~\ref{eq:gd} intractable. To address this, the proposed Latent Gradient Simulation Core parameterizes $\mathbf{\Phi}_{\hat{\theta}}(\cdot)$ and $\mathbf{\Phi}^{\top}_{\hat{\theta}}(\cdot)$ to simulate the role of $\mathbf{A}$ and $\mathbf{A}^{\dagger}$ during the gradient descent step under non-linear degradation. Leveraging the rich semantic and multi-scale information within the latent spaces from UNet blocks, the simulated gradient descent is performed in the latent spaces:
\begin{equation}
    \hat{\mathbf{f}}^{b, m} = \mathbf{f}_\mathbf{x}^{b, m} + \mathbf{\Phi}^{\top}_{\hat{\theta}}(\mathbf{f}_\mathbf{y}^{b, m}) - \mathbf{\Phi}^{\top}_{\hat{\theta}}(\mathbf{\Phi}_{\hat{\theta}}(\mathbf{f}_\mathbf{x}^{b, m})).
\label{eq:latent-unfold}
\end{equation}

\begin{figure}
    \centering
    \includegraphics[width=\linewidth]{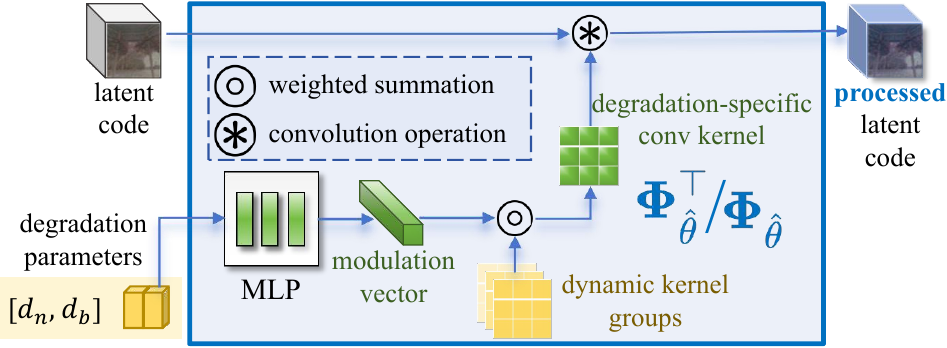}
    \caption{Details of $\mathbf{\Phi}_{\hat{\theta}}(\cdot)$ and $\mathbf{\Phi}^{\top}_{\hat{\theta}}(\cdot)$, both designed as 3$\times$3 degradation-aware dynamic convolutions. Despite having identical network structures, they do not share learned weights.}
    \label{fig:conv}
    \vspace{-4mm}
\end{figure}

In the context of the image linear inverse problem, $\mathbf{A}$ and $\mathbf{A}^{\dagger}$ are used for calculating the gradient direction. Specifically, in the SR task, they are typically modeled using hand-crafted parametrized convolution kernels \cite{wang2022ddnm}. Inspired by this, the proposed $\mathbf{\Phi}_{\hat{\theta}}(\cdot)$ and $\mathbf{\Phi}^{\top}_{\hat{\theta}}(\cdot)$ are designed to learn estimating different degradation types by using 3$\times$3 dynamic convolution  \cite{li2022odconv}, where $\hat{\theta}$ includes parameters of MLP-based dynamic weight generator and the convolution kernel groups which are dynamically assembled based on degradation parameters $\mathbf{d} =[d_n, d_b]$, as illustrated in Fig.~\ref{fig:conv}.
Via this way, the one-step generation process can incorporate degradation information as guidance, resulting in SR results that balance fidelity and realness. The pseudo-code of the LaGAR is shown in Algo.~\ref{alg:LaGAR}.

\begin{table*}[b]
\caption{Quantitative comparison of generative SR models on ODI-SR and SUN 360 datasets. ``model-Gen'' denotes models trained with a generative loss. ``Assessor360-m'' and ``Assessor360-o'' indicate variants \cite{wu2023assessor360} trained on the MVAQD \cite{jiang2021MVAQD} and OIQA \cite{duan2018OIQA} datasets, respectively. The best and second-best results are highlighted in
\boldred{red} and \boldblue{blue}.}
\centering
\resizebox{\linewidth}{!}{
\begin{tabular}{@{}c|c|ccccccc@{}}
\toprule
Datasets                   & Methods             & WS-PSNR↑ & WS-SSIM↑  & LPIPS↓ & DISTS↓  & FID↓   & Assesor360-m↑  & Assesor360-o↑  \\ \hline
\multirow{8}{*}{ODI-SR} 
% & Bicubic       & 23.26 & 0.5726 & \textbf{\textcolor{blue}{0.3113}} & 0.2048 & \textbf{\textcolor{red}{24.44}}  & 4.7581 & 65.92  & 0.6192  & 0.6771   \\
& S3Diff~\cite{2024s3diff}       & 20.91 & 0.4606  & 0.4197 & 0.1649  & 85.01   & \boldblue{0.6006}  & \boldblue{0.8049}    \\
& SeeSR~\cite{wu2024seesr}        & 22.14 & 0.5564  & 0.3775 & 0.1557  & 72.36   & 0.5795  & 0.7833   \\
% & SeeSR(bad)~\cite{wu2024seesr}        & 19.89 & 0.5629  & 0.4411 & 0.2234  & 126.21   & 4.9682  & 64.94 & \boldblue{0.4884} & \boldred{0.7141}   \\
& StableSR~\cite{StableSR_Wang_Yue_Zhou_Chan_Loy_2023}          & 22.29 & 0.5559  & 0.4359 & 0.1804  & 89.48   & 0.3419  & 0.6043    \\
& OSEDiff~\cite{wu2024osediff}        & 21.54 & 0.5545  & 0.3394 & 0.1559  & 83.27  & 0.5688  & 0.7804    \\
% & RealESR-GAN       & WS-PSNR↑ & WS-SSIM↑  & LPIPS↓ & DISTS↓  & FID↓   & NIQE↓  & MUSIQ↑ & MANIQA↑ & CLIPIQA↑  \\
& OSRT-Gen~\cite{osrt_Yu_Wang_Cao_Li_Shan_Dong_2023}      & 22.09 & 0.5598  & \boldblue{0.3129} & \boldblue{0.1439}  & 63.62   & 0.5400  & 0.7433   \\
& BPOSR-Gen~\cite{wang2024bposr}       & \boldred{23.41} & \boldred{0.6138}  & 0.3426 & 0.1666  & \boldblue{62.61}   & 0.4580  & 0.7347   \\
& OmniSSR~\cite{li2024omnissr}          & \boldblue{22.53} & 0.5263  & 0.6638 & 0.2747  & 113.79   & -0.0057  & 0.1527  \\
& \textbf{RealOSR (Ours)}     & 22.30 & \boldblue{0.5829}  & \boldred{0.2628} & \boldred{0.1194}  & \boldred{43.39}   & \boldred{0.6383}  & \boldred{0.8158}  \\ \hline
% & \textbf{RealOSR (Ours)}          & 23.72 & \textbf{\textcolor{blue}{0.6108}} & \textbf{\textcolor{red}{0.2941}} & \textbf{\textcolor{blue}{0.1976}} & 26.32  & 4.7097 & 67.97  & 0.6148  & 0.6683   \\ \hline
\multirow{8}{*}{SUN 360}   
% & Bicubic       & 28.03 & 0.7536 & 0.3284 & \textbf{\textcolor{blue}{0.2269}} & 148.98 & 6.5239 & 58.51  & 0.5601  & 0.6356   \\
& S3Diff~\cite{2024s3diff}       & 21.45 & 0.5015  & 0.4343 & 0.1520  & 94.67   & 0.6049  & 0.8142    \\
& SeeSR~\cite{wu2024seesr}        & 22.38 & 0.5916  & 0.3930 & 0.1479  & 78.62   & \boldblue{0.6504}  & \boldblue{0.8380}    \\
% & SeeSR(bad)~\cite{wu2024seesr}        & 20.01 & 0.5933  & 0.4403 & 0.2164  & 125.78   & 5.3113  & \boldred{70.54}   \\
& StableSR~\cite{StableSR_Wang_Yue_Zhou_Chan_Loy_2023}          & 22.55 & 0.5804  & 0.4663 & 0.1780  & 105.85   & 0.3134  & 0.5694   \\
& OSEDiff~\cite{wu2024osediff}        & 21.99 & 0.5921  & 0.3692 & 0.1502  & 85.54  & 0.5677  & 0.7918   \\
% & RealESR-GAN      & WS-PSNR↑ & WS-SSIM↑  & LPIPS↓ & DISTS↓  & FID↓   & NIQE↓  & MUSIQ↑ & MANIQA↑ & CLIPIQA↑   \\
& OSRT-Gen~\cite{osrt_Yu_Wang_Cao_Li_Shan_Dong_2023}          & 22.56 & 0.5953  & \boldblue{0.3451} & \boldblue{0.1407}  & 64.65   & 0.5133  & 0.7386  \\
& BPOSR-Gen~\cite{wang2024bposr}       & \boldred{23.59} & \boldred{0.6376}& 0.3907 & 0.1685  & \boldblue{62.37}   & 0.3834 & 0.6833  \\
& OmniSSR~\cite{li2024omnissr}         & 22.39 & 0.5220  & 0.7029 & 0.2889  & 140.32   & -0.0067  & 0.1330  \\
& \textbf{RealOSR (Ours)}      & \boldblue{22.70} & \boldblue{0.6171}  & \boldred{0.2888} & \boldred{0.1047} & \boldred{41.69}   & \boldred{0.6776}  & \boldred{0.8485}   \\ 
\bottomrule
\end{tabular}
}
\label{tab:main}
% \vspace{-2mm}
\end{table*}

\begin{algorithm}[h]
    % \scriptsize
    \caption{UNet Denoising with LaGAR}
    \label{alg:LaGAR}
    \KwIn{$\mathbf{z}^{(m)}_{\text{LR}}$, $\boldsymbol{\epsilon}_{\theta}=\{\text{Block}_{1}, \text{Block}_{2},...,\text{Block}_{B}\}$, $\mathbf{x}_\text{LR}^{(m)}$, $\mathbf{\Phi}_{\hat{\theta}}$, $\mathbf{\Phi}^{\top}_{\hat{\theta}}$}
    \KwOut{Denoised latent code $\tilde{\mathbf{z}}^{(m)}$}
    $\hat{\mathbf{f}}_\mathbf{x}^{0, m} = \mathbf{z}^{(m)}$\;
    \For{$b=1$ \KwTo $B$}{
        $\mathbf{f}_\mathbf{x}^{b, m} = \text{Block}_{b}(\hat{\mathbf{f}}^{b-1, m})$\;
        \small{\Comment{Latent-Pixel Transcoding Bridge}}
        $\mathbf{x}^{b, m} = \mathcal{A}^\text{L2P}_b(\mathbf{f}^{b, m})$\;
        $\mathbf{f}_\mathbf{x}^{b, m} = \mathcal{A}^\text{P2L}_b(\mathbf{x}^{b, m})$\;
        $\mathbf{f}^{b, m}_\mathbf{y} = \mathcal{A}^\text{P2L}_b(\mathbf{x}^{(m)}_{\text{LR}})$\;
        \small{\Comment{Latent Gradient Simulation Core}}
        $\hat{\mathbf{f}}^{b, m} = \mathbf{f}_\mathbf{x}^{b, m} + \mathbf{\Phi}^{\top}_{\hat{\theta}}(\mathbf{f}_\mathbf{y}^{b, m}) - \mathbf{\Phi}^{\top}_{\hat{\theta}}(\mathbf{\Phi}_{\hat{\theta}}(\mathbf{f}_\mathbf{x}^{b, m}))$\;
    }
    $\tilde{\mathbf{z}}^{(m)} = \hat{\mathbf{f}}_\mathbf{x}^{B, m}$\;
    \Return $\tilde{\mathbf{z}}^{(m)}$\;
\end{algorithm}

\subsection{Training Details}
\label{subsec:training details}
In our proposed RealOSR, the VAE Decoder and the degradation predictor remain frozen, requiring no training. The SD UNet and VAE Encoder are trained with LoRA conditioned on degradation parameters, enabling dynamic adjustments to model parameters in response to specific degradation. All weights in the LaGAR undergo training.

Following S3Diff \cite{2024s3diff}, we employ a comprehensive loss function and use adversarial learning to further enhance the realness of the SR results. The total loss is defined as:
% \vspace{-2mm}
\begin{equation}
\mathcal{L}_{total} = \lambda_\text{rec}\mathcal{L}_\text{rec} + \lambda_\text{LPIPS}\mathcal{L}_{\text{LPIPS}} + \lambda_\text{GAN} \mathcal{L}_\text{GAN}.
\label{eq:total-loss}
\end{equation}
Here, a differentiable variant of L1 Loss, named Charbonnier Loss~\cite{LapSRN}, is employed as the pixel-level reconstruction loss $\mathcal{L}_{\text{rec}}$, complemented by LPIPS Loss $\mathcal{L}_{\text{LPIPS}}$ and GAN Loss $\mathcal{L}_{\text{GAN}}$ to reduce the data distribution distance between SR results and HR ODIs.

Compared to directly applying ERP-bicubic downsampling \cite{deng2022omnidirectional}, OSRT \cite{osrt_Yu_Wang_Cao_Li_Shan_Dong_2023} introduces fisheye-bicubic downsampling to more accurately simulate real-world degradation in originally captured fisheye-formatted images, accounting for degradation that typically results from limited sensor precision and density before conversion into other storage formats. However, the occurring degradation during the real-world capture of raw images is usually unknown and more complex than simple bicubic downsampling. Therefore, we apply the degradation pipeline proposed in Real-ESRGAN \cite{wang2021realesrgan} to the fisheye images and subsequently convert them into LR ERP images to simulate real-world degradation.

\begin{figure*}[!h]
	%\newlength-4mm
	%\setlength{-4mm}{-0.4cm}
	\scriptsize
	\centering
	\begin{tabular}{l}
		\hspace{-0.42cm}
		\begin{adjustbox}{valign=t}
			\begin{tabular}{c}
				\includegraphics[width=0.260\textwidth]{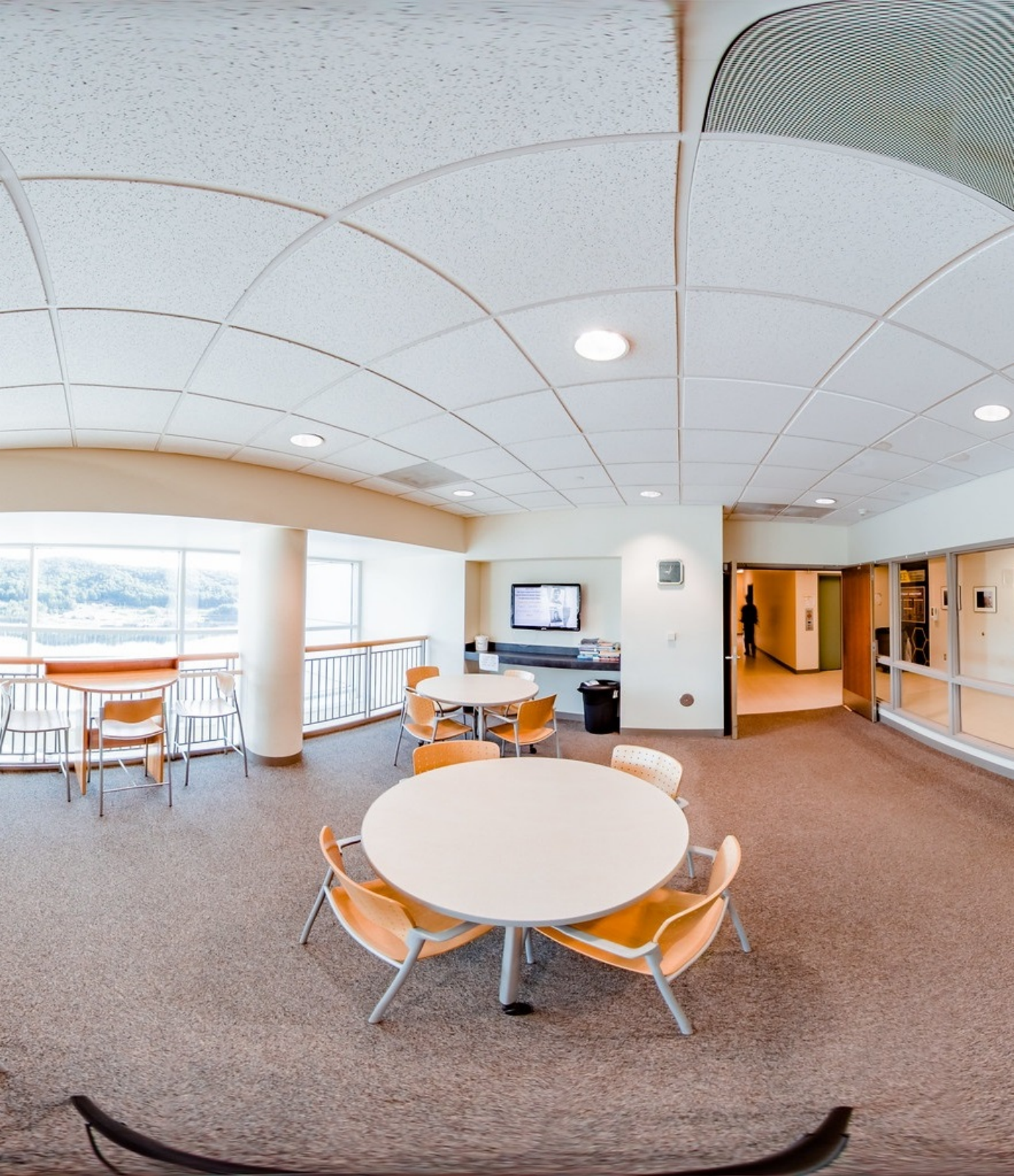}
				\\
				SUN 360: 0047
			\end{tabular}
		\end{adjustbox}
		\hspace{-2mm}
		\begin{adjustbox}{valign=t}
			\begin{tabular}{cccc}
				\includegraphics[width=0.149\textwidth]{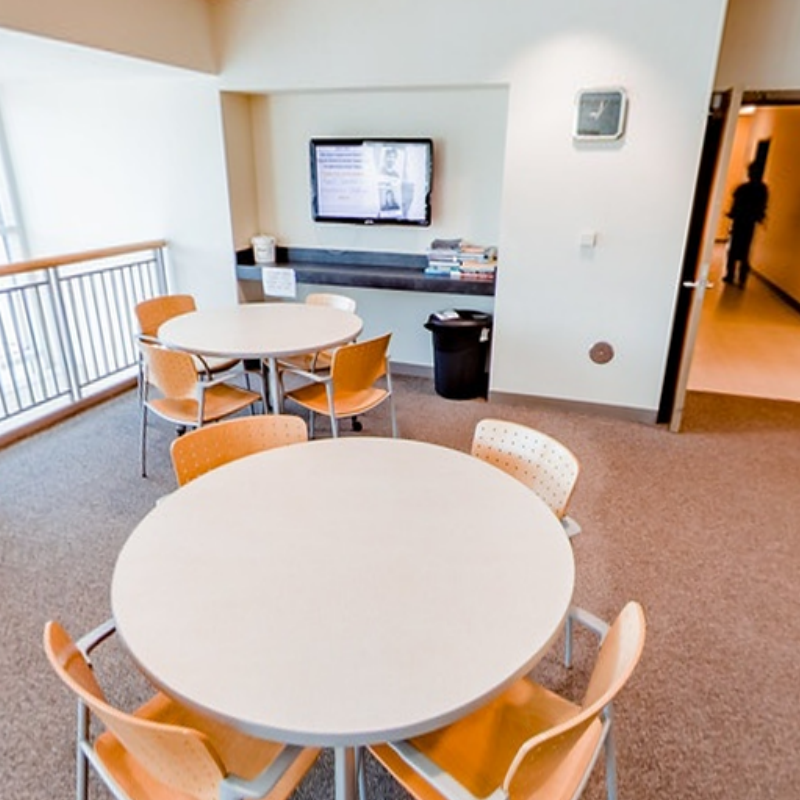} \hspace{-1mm} &
				\includegraphics[width=0.149\textwidth]{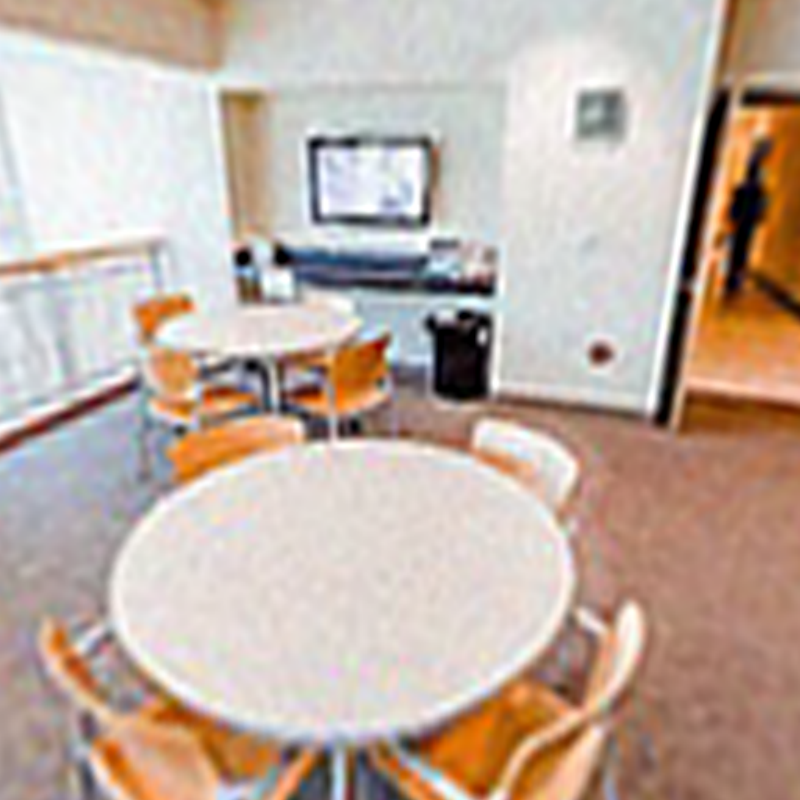} \hspace{-1mm} &
				\includegraphics[width=0.149\textwidth]{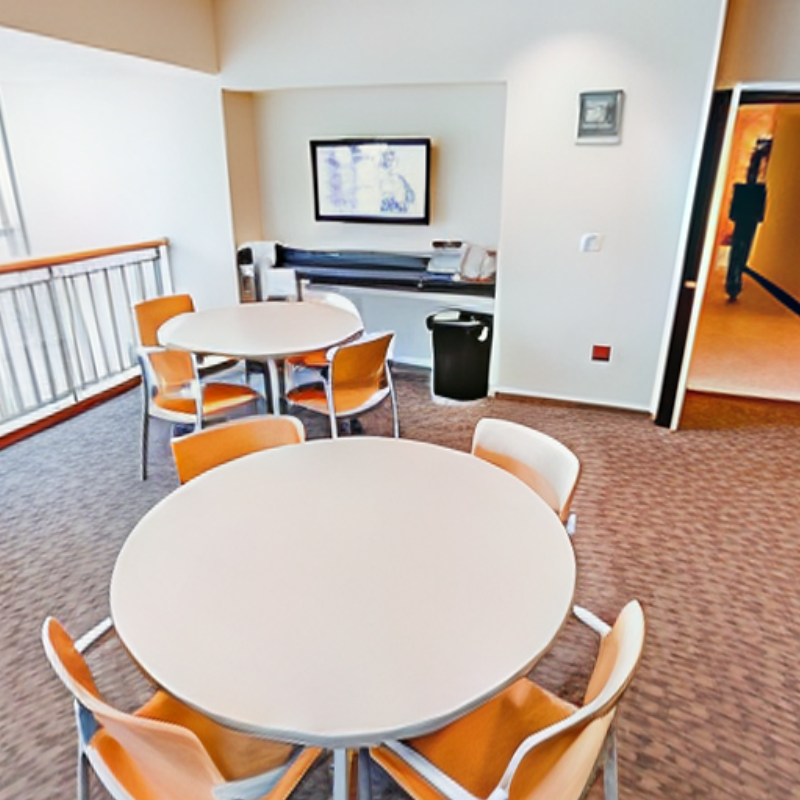} \hspace{-1mm} &
				\includegraphics[width=0.149\textwidth]{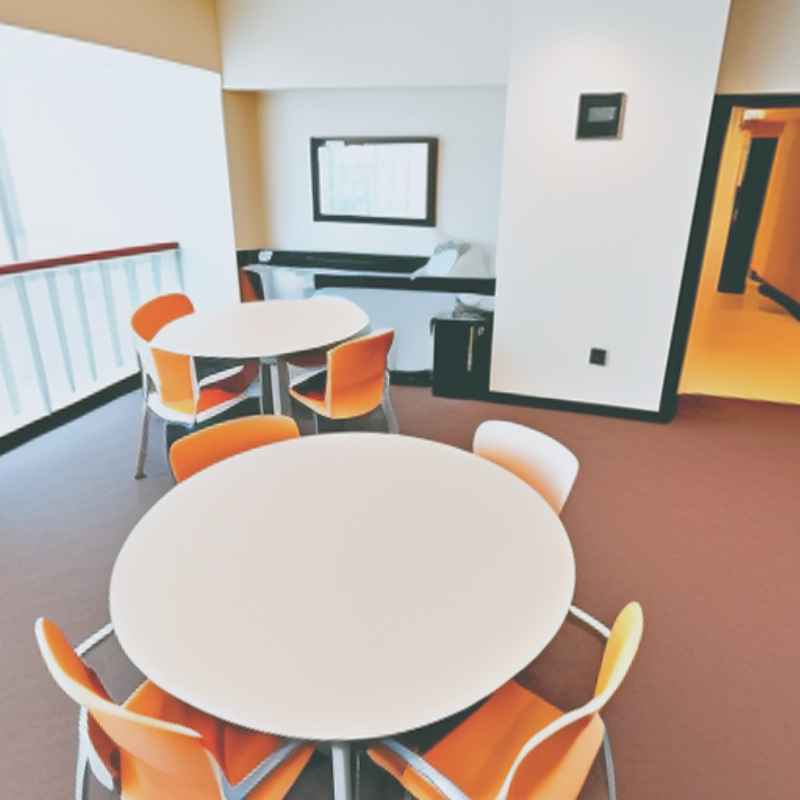} \hspace{-1mm} 
				\\
				GT \hspace{-1mm} &
				LR \hspace{-1mm} &
				S3Diff \cite{2024s3diff} \hspace{-1mm} &
				SeeSR \cite{wu2024seesr} \hspace{-1mm} 
				% \\ 
				% PSNR/SSIM \hspace{-1mm} &
				% 28.67dB/0.8317 \hspace{-1mm} &
				% 24.02dB/0.5849 \hspace{-1mm} &
				% 30.04dB/0.8855 \hspace{-1mm} 
				\\
				\includegraphics[width=0.149\textwidth]{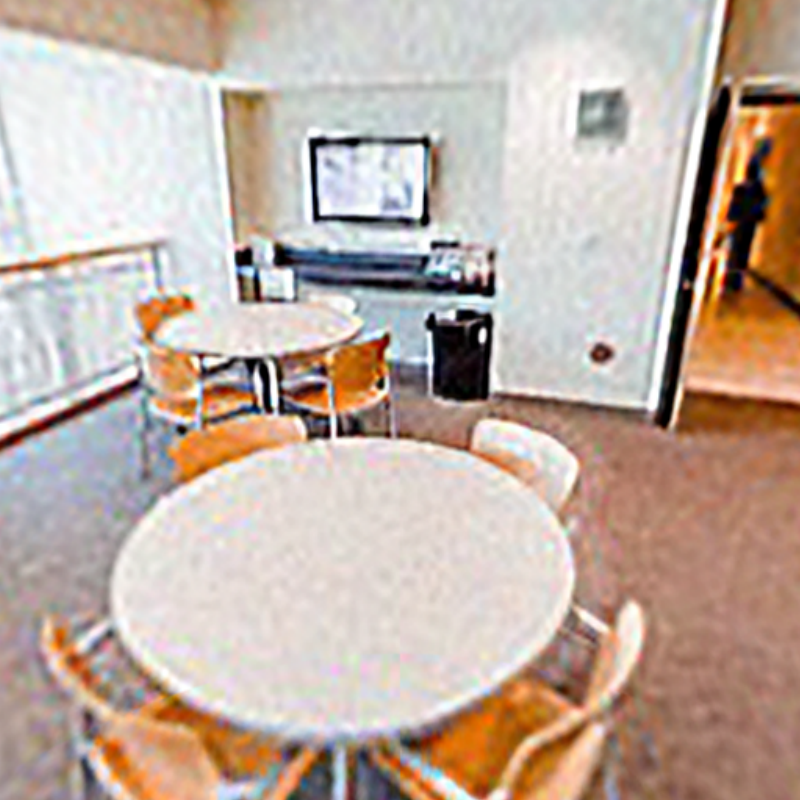} \hspace{-1mm} &
				\includegraphics[width=0.149\textwidth]{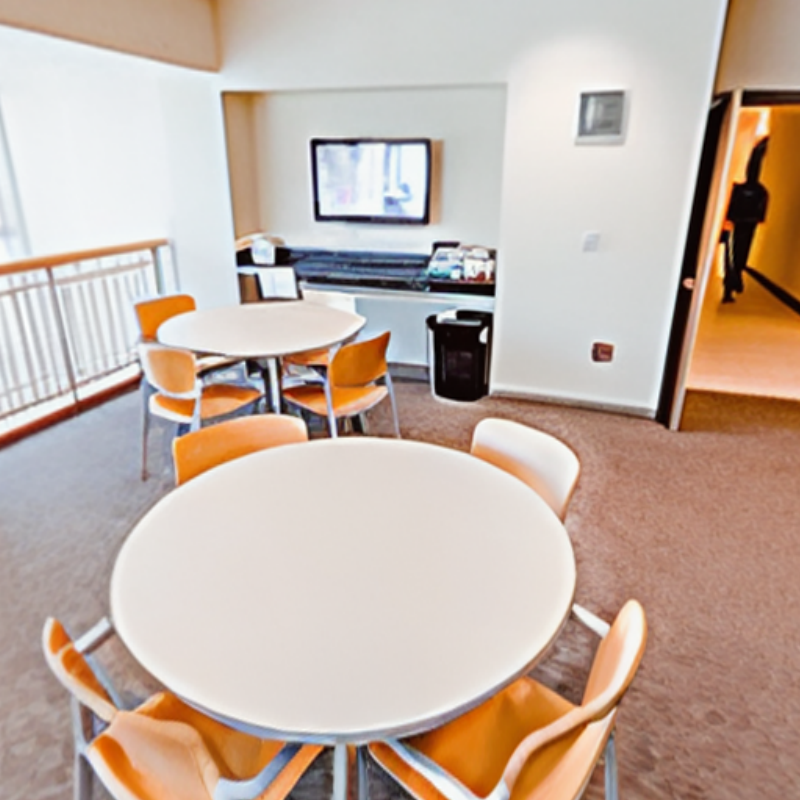} \hspace{-1mm} &
				\includegraphics[width=0.149\textwidth]{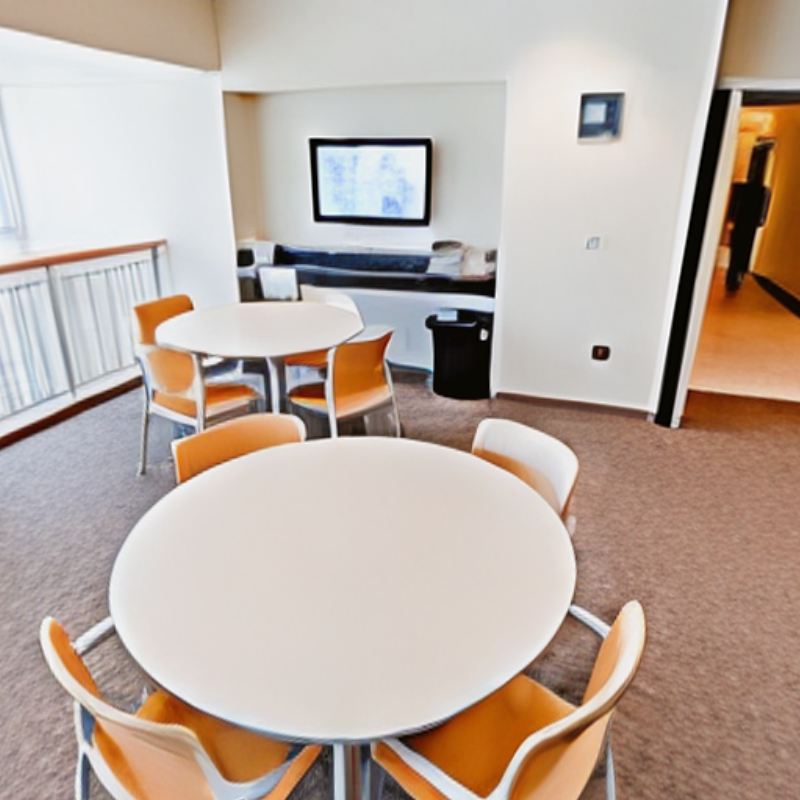} \hspace{-1mm} &
				\includegraphics[width=0.149\textwidth]{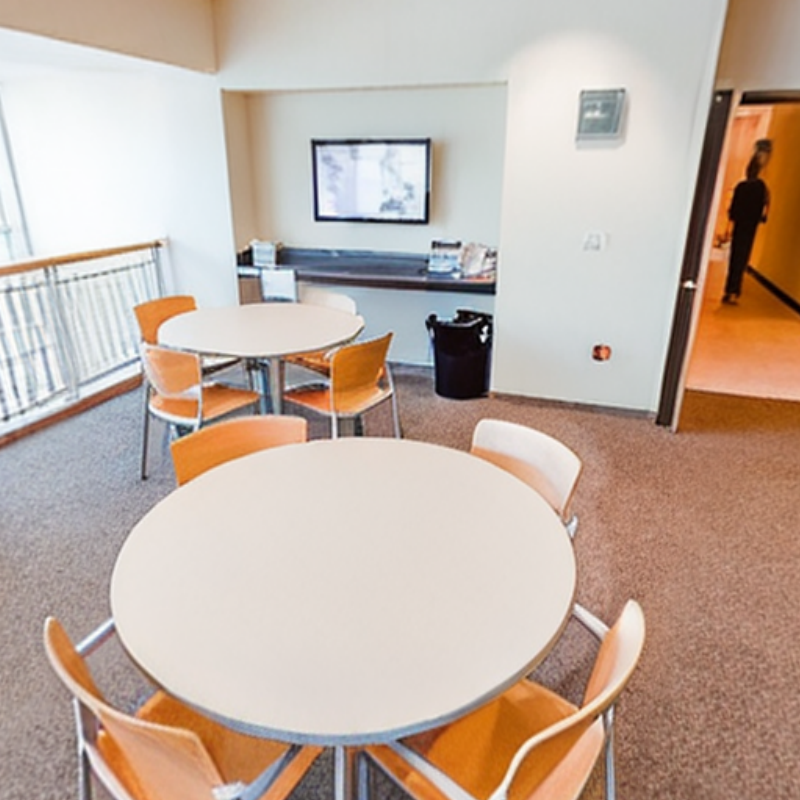} \hspace{-1mm}  
				\\ 
				OmniSSR \cite{li2024omnissr} \hspace{-1mm} &
				StableSR \cite{StableSR_Wang_Yue_Zhou_Chan_Loy_2023} \hspace{-1mm} &
				OSEDiff \cite{wu2024osediff} \hspace{-1mm} &
				\textbf{RealOSR} (ours) \hspace{-1mm} 
				% \\
				% 30.07dB/0.8802 \hspace{-1mm} &
				% 24.47dB/0.6421 \hspace{-1mm} &
				% 23.81dB/0.7384 \hspace{-1mm} &
				% 30.15dB/0.8859 \hspace{-1mm} 
			\end{tabular}
		\end{adjustbox}
		
		\\ 
		\hspace{-0.42cm}
		\begin{adjustbox}{valign=t}
			\begin{tabular}{c}
				\includegraphics[width=0.260\textwidth]{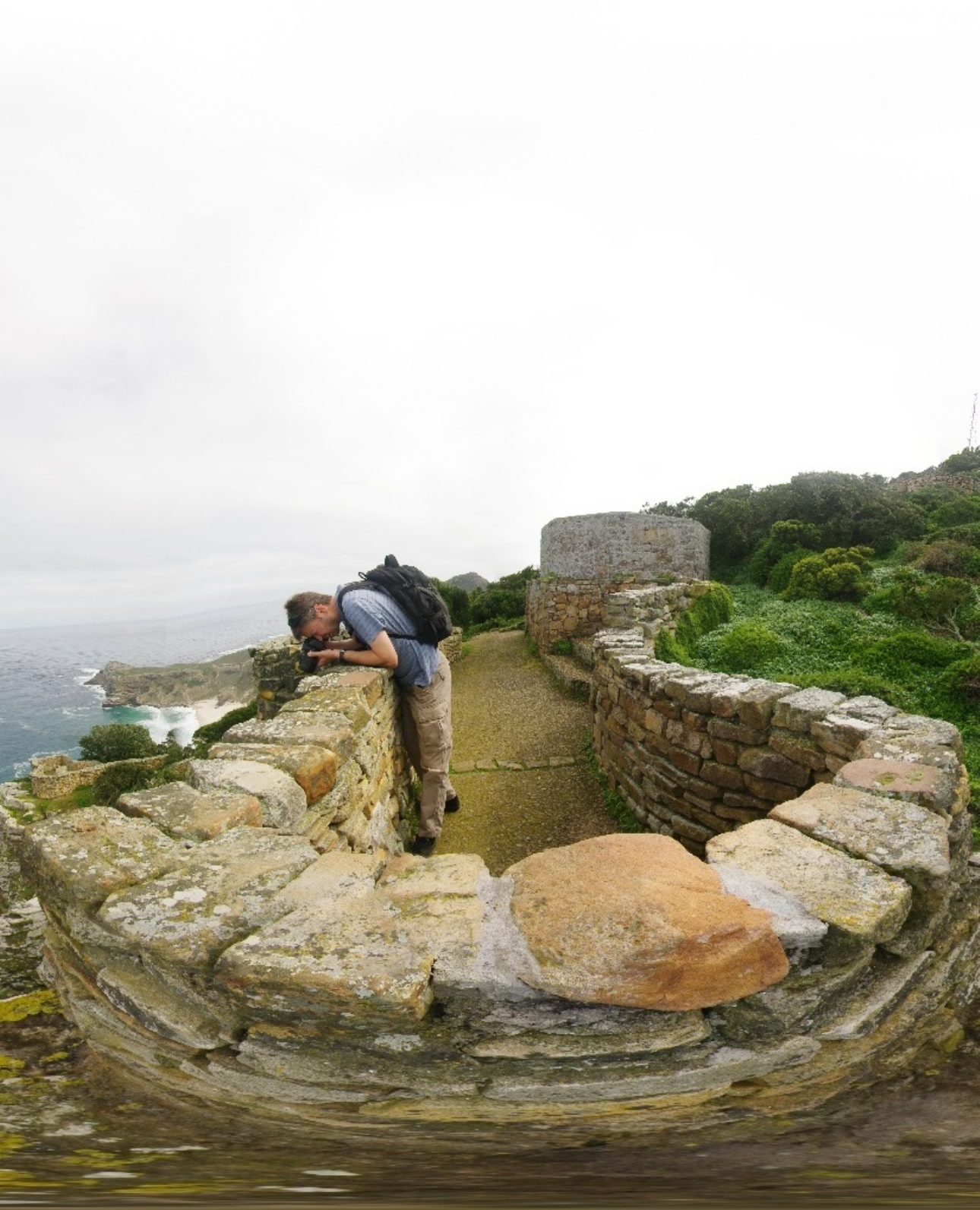}
				\\
				% SUN 360 : 047
                ODI-SR: 0095
			\end{tabular}
		\end{adjustbox}
		\hspace{-2mm}
		\begin{adjustbox}{valign=t}
			\begin{tabular}{cccc}
				\includegraphics[width=0.149\textwidth]{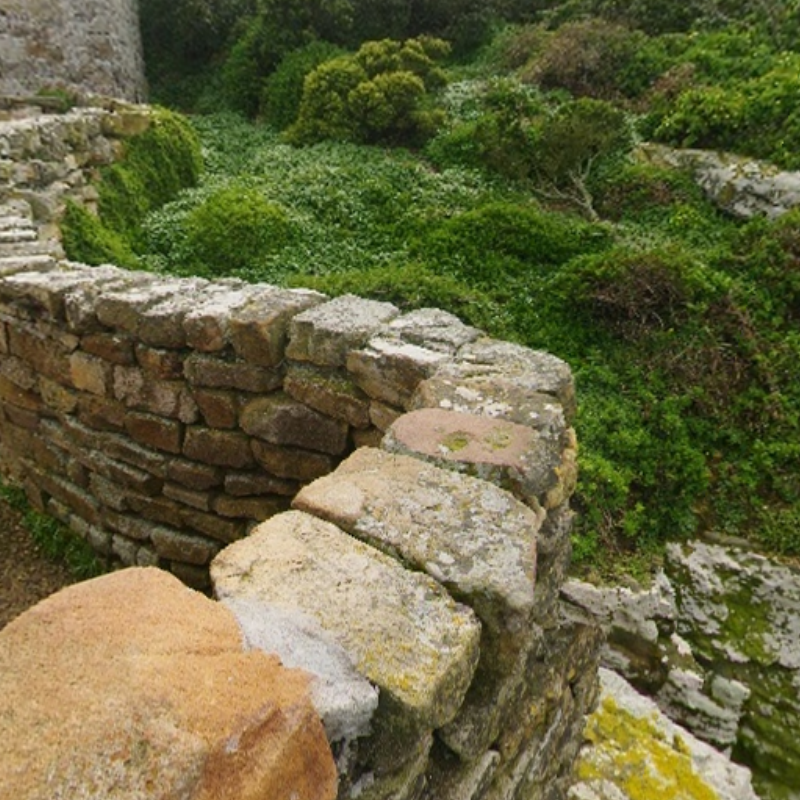} \hspace{-1mm} &
				\includegraphics[width=0.149\textwidth]{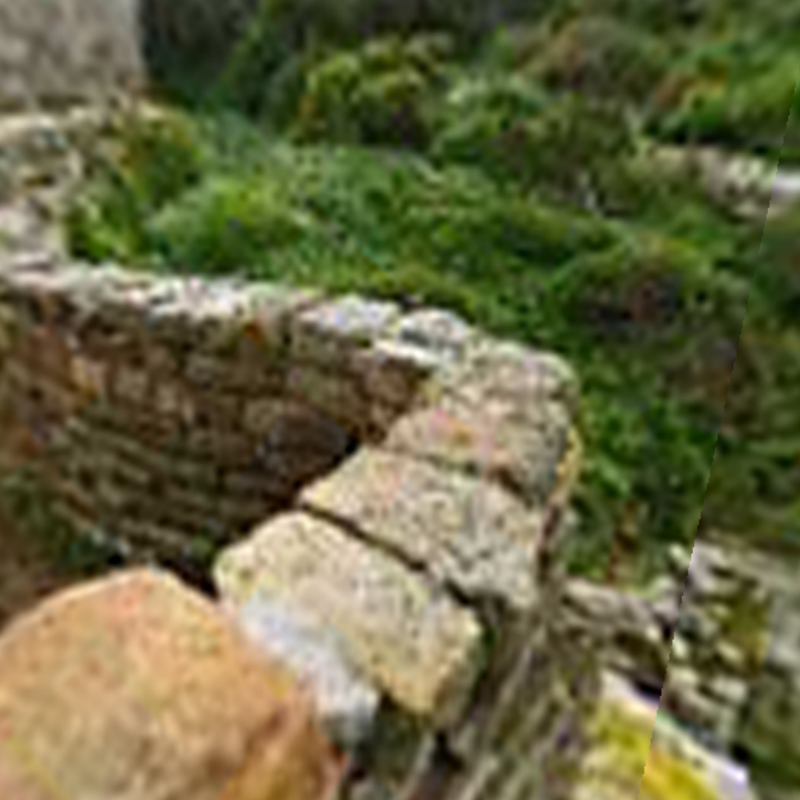} \hspace{-1mm} &
				\includegraphics[width=0.149\textwidth]{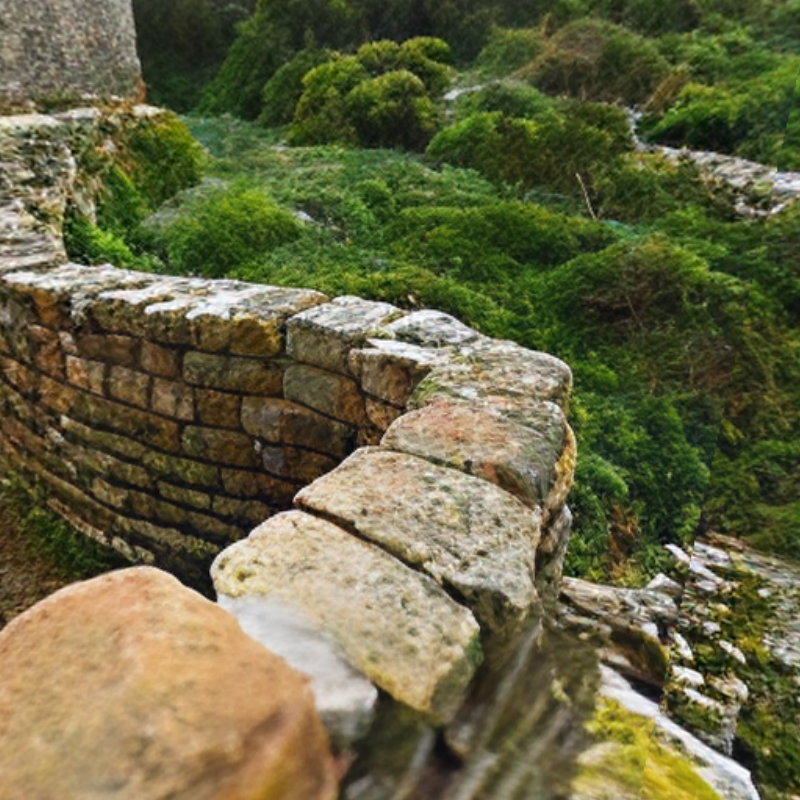} \hspace{-1mm} &
				\includegraphics[width=0.149\textwidth]{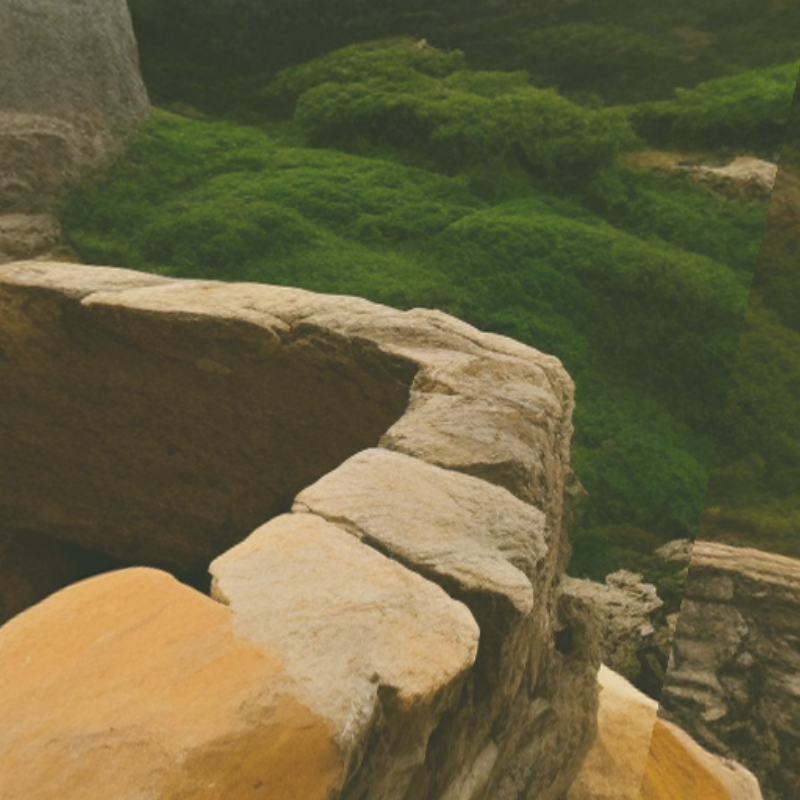} \hspace{-1mm} 
				\\
				GT \hspace{-1mm} &
				LR \hspace{-1mm} &
				S3Diff \cite{2024s3diff} \hspace{-1mm} &
				SeeSR \cite{wu2024seesr} \hspace{-1mm}  
				% \\
				% PSNR/SSIM \hspace{-1mm} &
				% 24.65dB/0.7753 \hspace{-1mm} &
				% 22.57dB/0.7188 \hspace{-1mm} &
				% 25.36dB/0.8029 \hspace{-1mm} 
				\\
				\includegraphics[width=0.149\textwidth]{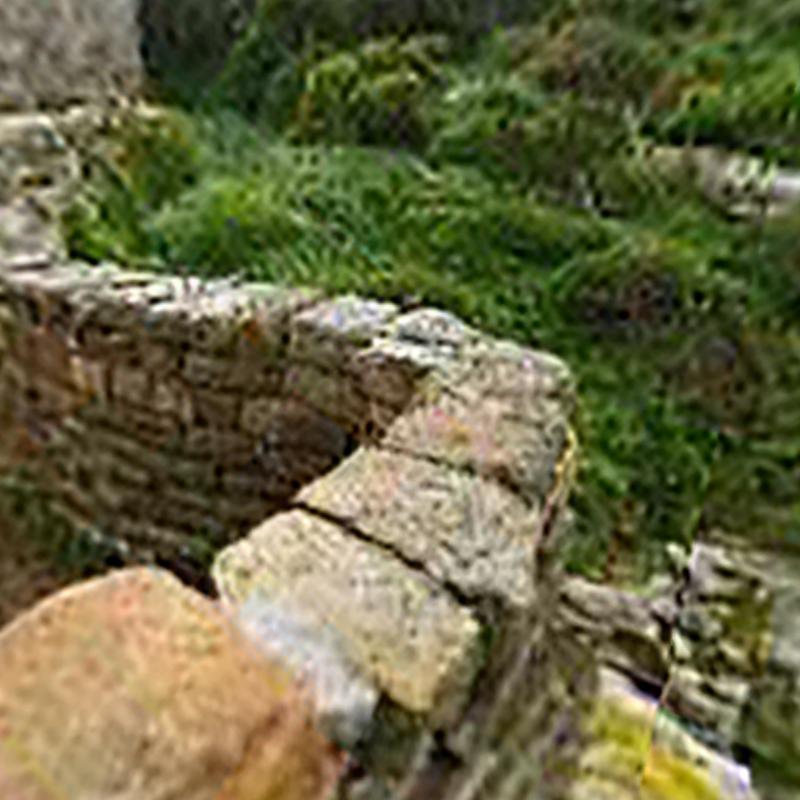} \hspace{-1mm} &
				\includegraphics[width=0.149\textwidth]{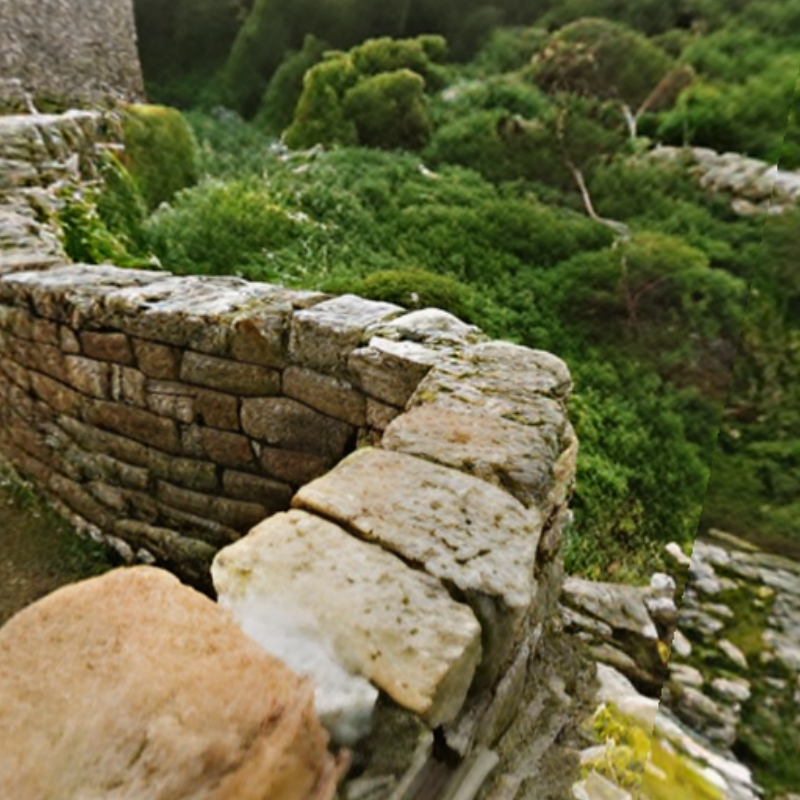} \hspace{-1mm} &
				\includegraphics[width=0.149\textwidth]{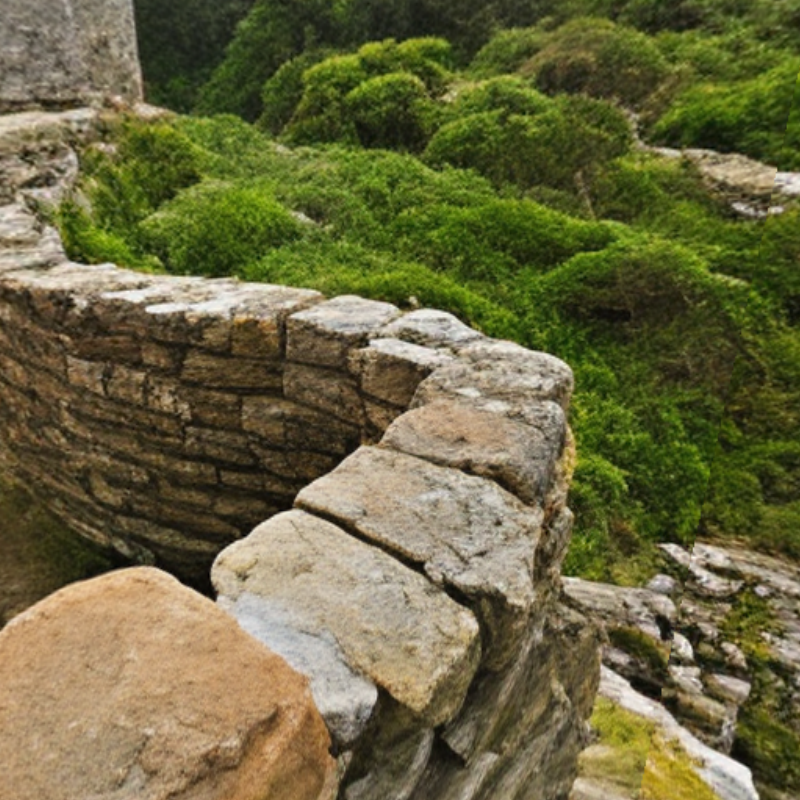} \hspace{-1mm} &
				\includegraphics[width=0.149\textwidth]{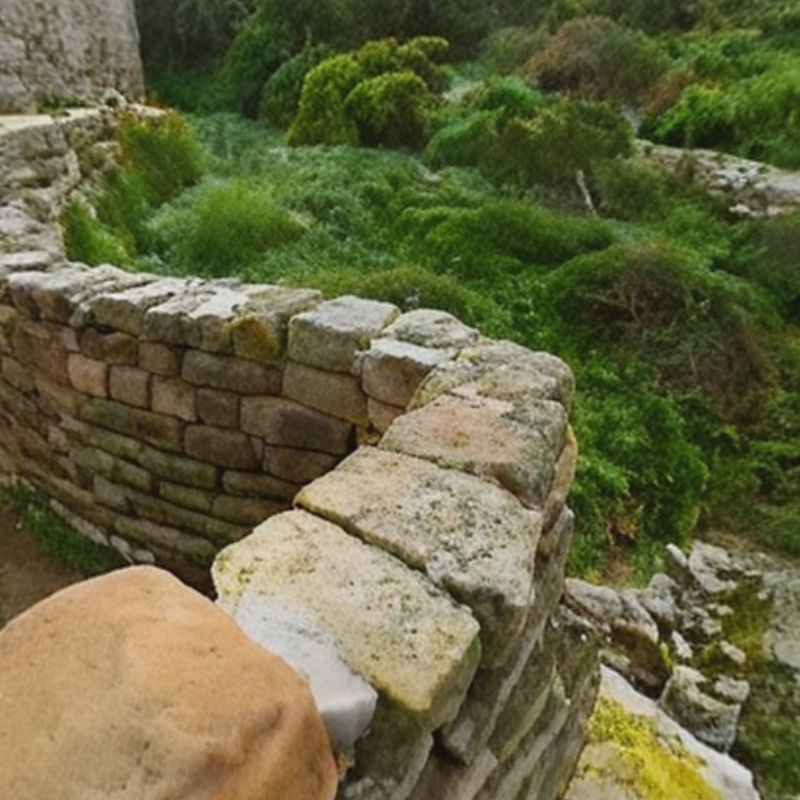} \hspace{-1mm}  
				\\ 
				OmniSSR \cite{li2024omnissr} \hspace{-1mm} &
				StableSR \cite{StableSR_Wang_Yue_Zhou_Chan_Loy_2023} \hspace{-1mm} &
				OSEDiff \cite{wu2024osediff} \hspace{-1mm} &
				\textbf{RealOSR} (ours) \hspace{-1mm}
				% \\
				% 23.02dB/0.7525 \hspace{-1mm} &
				% 23.48dB/0.7799 \hspace{-1mm} &
				% 24.62dB/0.7981 \hspace{-1mm} &
				% 26.53dB/0.8265 \hspace{-1mm} 
			\end{tabular}
		\end{adjustbox}
		\\ 
		
	\end{tabular}
	\caption{Visualized comparison of SR results on SUN 360 test set and ODI-SR test set. 0047 and 0095 are the ID numbers in the test set filenames. Our RealOSR can achieve photo-realistic SR results compared to other generative methods. Note that OmniSSR~\cite{li2024omnissr} assumes a linear and fully known degradation operator $\mathbf{A}$, which does not hold for real-world degradations that are typically nonlinear and unknown.}
    \vspace{-2mm}
	\label{fig:diffusion_visual}
\end{figure*}

\section{Experiments}
\label{sec:experiments}
% 实现细节
\subsection{Implementation Details}
\noindent \textbf{Datasets and Pretrained Models.} 
We use the ODI-SR dataset from LAU-Net \cite{Deng_Wang_Xu_Guo_Song_Yang_2021} and the SUN 360 Panorama dataset \cite{Jianxiong_Xiao_Ehinger_Oliva_Torralba_2012}, where ground truth (GT) ERP images are of resolution 1024$\times$2048. The $\times4$ downsampling degradation pipeline of Real-ESRGAN \cite{wang2021realesrgan} is performed on GT fisheye images and stored as LR ERP images. The version of Stable Diffusion is SD-Turbo~\cite{sd_turbo}.

% \noindent \textbf{Compared Methods.}
\noindent \textbf{Training settings.}
The hyperparameters $\lambda_\text{rec}$, $ \lambda_\text{LPIPS}$, $\lambda_\text{GAN}$ are set as 2, 5, and 0.5, respectively. Optimization is performed using Adam with a learning rate of 1e-5 and a total batch size of 4 for \textit{50k} iterations. All experiments run on NVIDIA 4090. Except for training BPOSR~\cite{wang2024bposr} on NVIDIA A6000, as its memory usage exceeds 24GB.

\noindent \textbf{Evaluation Metrics.} To comprehensively assess the performance of various methods, we employ both reference-based and non-reference metrics. WS-PSNR \cite{WS_PSNR_Sun_Lu_Yu_2017} and WS-SSIM \cite{WS_SSIM_Zhou_Yu_Ma_Shao_Jiang_2018} are reference-based fidelity metrics computed on RGB channels. LPIPS \cite{zhang2018perceptual} and DISTS \cite{dists_ding2020image} serve as reference-based perceptual quality metrics, while FID \cite{fid} measures the distribution distance between GT images and SR images. 
Common non-reference metrics (e.g., NIQE~\cite{niqe}, MANIQA~\cite{yang2022maniqa}, MUSIQ~\cite{musiq}, CLIPIQA~\cite{clipiqa}) are designed for planar images and fail to accurately reflect ODI quality. Thus, we adopt Assessor360~\cite{wu2023assessor360}, a non-reference ODI quality assessment model by simulating human exploration of ODIs through diverse viewport sequences. Assessor360 offers multiple checkpoints trained on different datasets. We use the checkpoints trained on MVAQD \cite{jiang2021MVAQD} and OIQA \cite{duan2018OIQA}, which focus on JPEG compression, noise, and blur, making them well aligned with Real-ESRGAN and real-world degradation.
% Additionally, NIQE \cite{niqe}, MANIQA \cite{yang2022maniqa}, MUSIQ \cite{musiq}, and CLIPIQA \cite{clipiqa} are non-reference perceptual quality metrics.

\subsection{Evaluation of RealOSR}
\label{subsec: main_exp}
This section evaluates RealOSR by categorizing the methods for comparison into two groups: 

(1) \textbf{Diffusion-based SR methods}: S3Diff~\cite{2024s3diff}, SeeSR~\cite{wu2024seesr}, StableSR~\cite{StableSR_Wang_Yue_Zhou_Chan_Loy_2023}, and OSEDiff~\cite{wu2024osediff} are recent diffusion-based PISR methods. OmniSSR~\cite{li2024omnissr} is a diffusion-based ODISR method. They are all generative models.

(2) \textbf{End-to-end SR methods}: OSRT~\cite{osrt_Yu_Wang_Cao_Li_Shan_Dong_2023}, BPOSR~\cite{wang2024bposr} are recent cutting-edge ODISR methods. \textbf{As most ODISR methods have no publicly available code}, we include the representative PISR method SwinIR~\cite{liang2021swinir} as a supplement. They are originally trained as regressive models with pixel-level reconstruction loss (e.g., $L_1$ loss), denoted as ``model-Reg" in Tab.~\ref{tab:pix-recon}. For fair comparison, we also train them as generative models with our generative loss in Eq.~\ref{eq:total-loss}, denoted as ``model-Gen" in Tab.~\ref{tab:main}.

\textbf{Comparison with generative SR methods} are shown in Tab.~\ref{tab:main} and Fig.~\ref{fig:diffusion_visual}. RealOSR demonstrates a marked improvement in both fidelity and realness compared to existing generative SR methods, particularly in preserving color consistency and fine details aligned with the input LR images.
In Fig.~\ref{fig:diffusion_visual}, RealOSR produces results more faithful to the ground truth, notably in the fine floor textures of (SUN 360: 0047) and the detailed reconstruction of rocky surfaces in (ODI-SR: 0095). Moreover, for (ODI-SR: 0095), S3Diff~\cite{2024s3diff} and OSEDiff~\cite{wu2024osediff} exhibit noticeable color shifts in the rocky and grassy regions, whereas RealOSR maintains accurate and consistent color fidelity.

\begin{figure}[!h]
  \centering
  % \begin{minipage}{0.50\textwidth}
    % \centering
    \includegraphics[width=\linewidth]{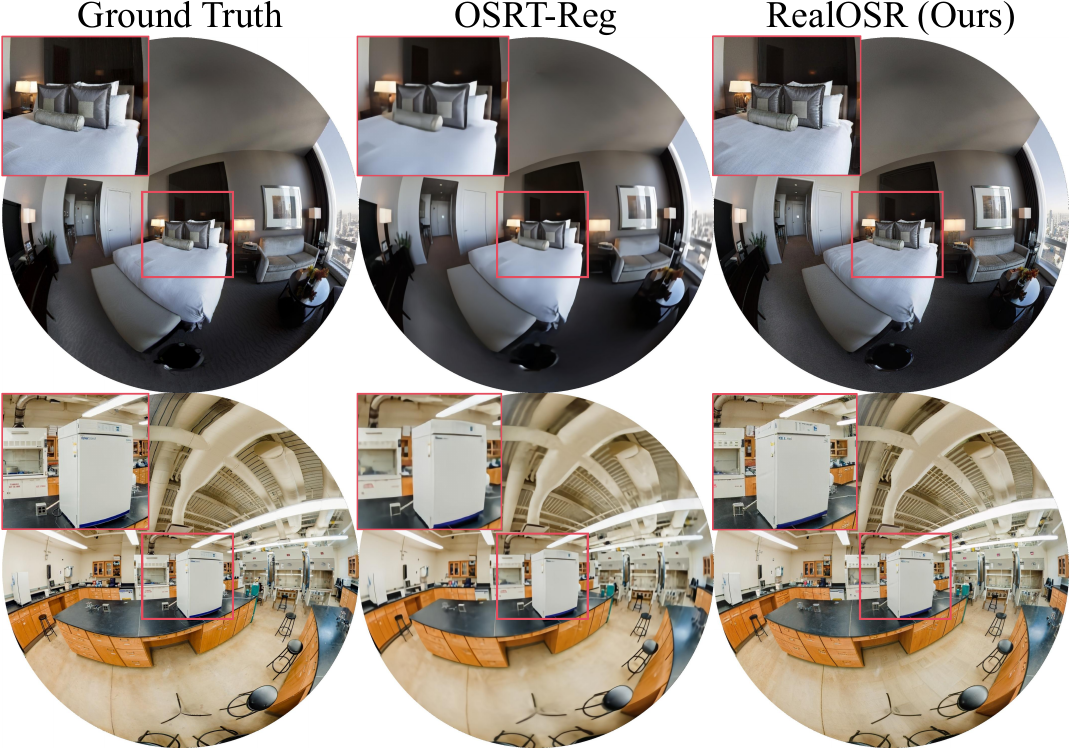}
    \caption{Comparison between OSRT-Reg and our RealOSR displayed in fisheye projection. Zoom in for more details.}
    \label{fig:fisheye}
    \vspace{-10pt}
  % \end{minipage}
\end{figure}

\textbf{Comparison with regressive SR methods} are shown in Tab.~\ref{tab:pix-recon} and the visual results are presented in Fig.~\ref{fig:fisheye}. Our method has superior photo-realistic performance compared to end-to-end trained regressive methods, avoiding over-smooth and distorted visual content.
In Fig.~\ref{fig:fisheye}, RealOSR preserves the intricate textures of the pillows with high fidelity, while OSRT-Reg~\cite{osrt_Yu_Wang_Cao_Li_Shan_Dong_2023} produces an over-smoothed result lacking fine structural detail.

\begin{table*}[h]
    \caption{Quantitative comparison of regressive SR models on ODI-SR and SUN 360 datasets. ``model-Reg'' denotes models trained with a regressive loss. ``Assessor360-m'' and ``Assessor360-o'' indicate variants \cite{wu2023assessor360} trained on the MVAQD \cite{jiang2021MVAQD} and OIQA \cite{duan2018OIQA} datasets, respectively. The best and second-best results are highlighted in \boldred{red} and \boldblue{blue}.}

    \centering
\resizebox{\linewidth}{!}{
\begin{tabular}{@{}c|c|ccccccccc@{}}
\toprule
Datasets                   & Methods             & WS-PSNR↑ & WS-SSIM↑  & LPIPS↓ & DISTS↓  & FID↓   & Assesor360-m↑  & Assesor360-o↑ \\ \hline
\multirow{4}{*}{ODI-SR} 
% & Bicubic       & 23.26 & 0.5726 & \textbf{\textcolor{blue}{0.3113}} & 0.2048 & \textbf{\textcolor{red}{24.44}}  & 4.7581 & 65.92  & 0.6192  & 0.6771   \\
& SwinIR-Reg~\cite{liang2021swinir}        & \boldblue{24.23} & \boldblue{0.6546}  & 0.4887 & 0.2695  & 88.92   & 0.1320  & 0.2674    \\
& OSRT-Reg~\cite{osrt_Yu_Wang_Cao_Li_Shan_Dong_2023}       & \boldred{24.36} & \boldred{0.6589}  & \boldblue{0.4855} & \boldblue{0.2679}  & \boldblue{84.60}   & \boldblue{0.1374}  & \boldblue{0.2762}   \\
& BPOSR-Reg~\cite{wang2024bposr}       & 24.20 & 0.6519  & 0.4995 & 0.2778  & 95.04   & 0.1102  & 0.2465   \\
& \textbf{RealOSR (Ours)}          & 22.30 & 0.5829  & \boldred{0.2628} & \boldred{0.1194}  & \boldred{43.39}   & \boldred{0.6383}  & \boldred{0.8158}  \\ \hline
\multirow{4}{*}{SUN 360}

& SwinIR-Reg~\cite{liang2021swinir}        & \boldblue{24.35} & \boldblue{0.6808}  & 0.4977 & 0.2578  & 88.17   & 0.1987  & 0.3198   \\
& OSRT-Reg~\cite{osrt_Yu_Wang_Cao_Li_Shan_Dong_2023}       & \boldred{24.55} & \boldred{0.6862}  & \boldblue{0.4841} & \boldblue{0.2538}  & \boldblue{82.00}   & \boldblue{0.2276}  & \boldblue{0.3458}   \\
& BPOSR-Reg~\cite{wang2024bposr}       & 24.30 & 0.6766  & 0.5105 & 0.2657  & 90.95   & 0.1597  & 0.2885  \\
& \textbf{RealOSR (Ours)}      & 22.70 & 0.6171  & \boldred{0.2888} & \boldred{0.1047} & \boldred{41.69}   & \boldred{0.6776}  & \boldred{0.8485} \\ \bottomrule
\end{tabular}
}
    \label{tab:pix-recon}
    % \vspace{-4mm}
\end{table*}

\subsection{Ablation Study}
This section investigates the contributions of the two components in LaGAR, \textbf{Latent-Pixel Transcoding Bridge} and \textbf{Latent Gradient Simulation Core}. 

The following model variants are considered: \ding{172} RealOSR (No-Guide): Remove LaGAR from RealOSR, only with TP images as input; \ding{173} RealOSR (Latent-Add): Latent Gradient Simulation Core is replaced by directly adding $\mathbf{f}^{b,m}_\mathbf{y}$ and $\mathbf{f}^{b,m}_\mathbf{x}$; \ding{174} RealOSR (Pixel-Guide): Latent Gradient Simulation Core is applied in the pixel space; \ding{175} RealOSR: The full proposed model.

The results in Tab.~\ref{tab:ablation} show that:
(1) Comparing \ding{172} and \ding{173} demonstrates that Latent-Pixel Transcoding Bridge facilitates efficient alignment and conversion of LR image information, leading to certain performance improvements.
(2) Comparing \ding{173} and \ding{175} highlights that Latent Gradient Simulation Core provides more effective degradation guidance compared to simply adding information from the LR image.
(3) Comparing \ding{174} and \ding{175} reveals that gradient simulation in the latent space enables better modeling of real-world degradation guidance.

\begin{table}[!h]
        % \vspace{-3mm}
        \caption{Ablation of LaGAR, best results shown in \boldred{red}.}
        \centering
        % \resizebox{0.5\linewidth}{!}{
        \begin{tabular}{l|cccc}
        \toprule
            ID & Method & LPIPS$\downarrow$ & DISTS$\downarrow$ & FID$\downarrow$ \\
            \hline
            \ding{172} &  RealOSR (No-Guide)  & 0.2763 & 0.1203 & 49.78\\
            \ding{173} &  RealOSR (Latent-Add)  & 0.2686 & 0.1246 & 45.81\\
            \ding{174} &  RealOSR (Pixel-Guide)  & 0.2669 & 0.1252 & 46.35\\
            \ding{175} &  RealOSR (\textbf{Ours})  & \boldred{0.2628} & \boldred{0.1194} & \boldred{43.39}\\
        \bottomrule
        \end{tabular}
        % }
    \label{tab:ablation}
    \vspace{-8pt}
\end{table}

\begin{table}[h]
    \caption{Efficiency comparison with diffusion-based and end-to-end models, in parameter count (1M = 1$\times10^6$) and inference time per ERP image. ``RealOSR-s'' and ``RealOSR-p'' denote the \textbf{serial} and \textbf{parallel} super-resolution of all TP images, respectively. The best and second-best results among diffusion-based models are shown in \boldred{red} and \boldblue{blue}.}
    \centering
    \begin{tabular}{l|c|c}
    \toprule
        Method & Inference Time (s) & \#Params (M)  \\
        \hline
        S3Diff~\cite{2024s3diff}  & 10.42 & \boldred{1326.76} \\
        SeeSR~\cite{wu2024seesr}  & 42.32 & 2249.91 \\
        OmniSSR~\cite{li2024omnissr} & 511.70 & 1554.64 \\
        % OmniSSR & 511.70 & 1554.64 \\
        StableSR~\cite{StableSR_Wang_Yue_Zhou_Chan_Loy_2023}  & 162.96 & 1554.64 \\
        RealOSR-s  & \boldblue{6.85} & \boldblue{1352.20} \\
        RealOSR-p  & \boldred{2.36} & \boldblue{1352.20} \\
        % \hline
        % OSRT~\cite{osrt_Yu_Wang_Cao_Li_Shan_Dong_2023}  & 3.52 & 11.93 \\
        % BPOSR~\cite{wang2024bposr}  & 0.25 & 2.19 \\
    \bottomrule
    \end{tabular}
    \vspace{-4mm}
    \label{tab:efficiency}
\end{table}

\subsection{Discussion}
\label{sec:discussion}

\textbf{Efficiency Comparison.} 
RealOSR achieves superior efficiency compared to other diffusion-based methods due to its unique design under the one-step denoising framework (see Tab.~\ref{tab:efficiency}). It achieves a \textbf{200$\times$} speed-up over the latest diffusion-based ODISR method, OmniSSR. Although the LaGAR modules introduce 25.44M additional parameters, RealOSR still surpasses S3Diff in efficiency. Furthermore, by performing parallel SR on multiple TP images, RealOSR achieves an inference time of \textbf{2.36 seconds}.
Given the high resolution of ERP images, direct processing in a single pass is infeasible. Diffusion-based PISR methods adopt a shifting-window strategy for ERP images, leading to redundant computations. In contrast, RealOSR employs TP images for more efficient and compact partitioning. 
A detailed speed-up analysis is provided in \textbf{Sec.~II in the supplementary materials}.

\textbf{Robustness Comparison.} 
We assess the robustness of various methods under more severe degradation conditions. This involves increasing JPEG compression and random noise levels in the input images. As shown in Tab.~\ref{tab:severe_degrade}, RealOSR exhibits \textbf{superior} and \textbf{consistent} SR performance (FID: 43.39 $\rightarrow$ 49.31) under more severe degradation, significantly outperforming other approaches such as S3Diff (FID: 85.01 $\rightarrow$ 124.14). This highlights the robustness of our approach in handling varying levels of degradation encountered in real-world scenarios.

Additionally, RealOSR exhibits strong cross-domain robustness in low-light night scenes. To verify this, we generated low-light GT by reducing the Y-channel values of the original GT images, and subsequently degraded them to produce low-light LR images. As shown in Tab.~\ref{tab:low_light}, RealOSR consistently achieves superior results across all metrics, underscoring its cross-domain robustness.

\begin{table}[h]
\caption{Quantitative comparison of diffusion-based methods \textbf{under more severe degradation conditions}. The best and second-best results are highlighted in \boldred{red} and \boldblue{blue}.}
\centering
\resizebox{\linewidth}{!}{
\begin{tabular}{@{}c|c|ccccc@{}}
\toprule
Datasets                   & Methods             & WS-PSNR↑ & WS-SSIM↑  & LPIPS↓ & DISTS↓  & FID↓    \\ \hline
% & RealOSR (Ours)          & 23.72 & \textbf{\textcolor{blue}{0.6108}} & \textbf{\textcolor{red}{0.2941}} & \textbf{\textcolor{blue}{0.1976}} & 26.32  & 4.7097 & 67.97  & 0.6148  & 0.6683   \\ \hline
\multirow{5}{*}{SUN 360}   
% & Bicubic       & 28.03 & 0.7536 & 0.3284 & \textbf{\textcolor{blue}{0.2269}} & 148.98 & 6.5239 & 58.51  & 0.5601  & 0.6356   \\
& S3Diff       & 20.42 & 0.4066  & 0.5194 & 0.1903  & 124.14    \\

& OmniSSR         & 20.59 & 0.3975  & 0.7945 & 0.3312  & 188.04   \\

& StableSR          & \boldblue{21.36} & 0.4821  & 0.5526 & 0.2138  & 139.79  \\

& OSEDiff        & \boldblue{21.36} & \boldblue{0.5300}  & \boldblue{0.4397} & \boldblue{0.1775}  & \boldblue{112.36}  \\

& RealOSR (Ours)      & \boldred{22.25} & \boldred{0.5967}  & \boldred{0.3155} & \boldred{0.1096}  & \boldred{49.31}  \\ \bottomrule
% & RealOSR (Ours)      & \boldred{22.24} & \boldred{0.5948}  & \boldred{0.3131} & \boldred{0.1128}  & \boldred{51.63}   & 3.4158  & \boldblue{69.26} & 0.4219 & \boldblue{0.6065}  \\ \bottomrule
\end{tabular}
}
\label{tab:severe_degrade}
\vspace{-2mm}
\end{table}

\begin{table}[h]
\caption{Quantitative comparison of diffusion-based methods \textbf{in low-light night scenes}. The best and second-best results are highlighted in \boldred{red} and \boldblue{blue}.}
\centering
\resizebox{\linewidth}{!}{
\begin{tabular}{@{}c|c|ccccc@{}}
\toprule
Datasets                   & Methods             & WS-PSNR↑ & WS-SSIM↑  & LPIPS↓ & DISTS↓  & FID↓    \\ \hline
% & RealOSR (Ours)          & 23.72 & \textbf{\textcolor{blue}{0.6108}} & \textbf{\textcolor{red}{0.2941}} & \textbf{\textcolor{blue}{0.1976}} & 26.32  & 4.7097 & 67.97  & 0.6148  & 0.6683   \\ \hline
\multirow{3}{*}{SUN 360}   
% & Bicubic       & 28.03 & 0.7536 & 0.3284 & \textbf{\textcolor{blue}{0.2269}} & 148.98 & 6.5239 & 58.51  & 0.5601  & 0.6356   \\
& S3Diff       & 26.49 & 0.5171  & 0.5346 & 0.2423  & 179.46    \\

& OSEDiff        & \boldblue{27.89} & \boldblue{0.6069}  & \boldblue{0.4545} & \boldblue{0.2269}  & \boldblue{163.69}  \\

& RealOSR (Ours)      & \boldred{29.46} & \boldred{0.7293}  & \boldred{0.3068} & \boldred{0.1387}  & \boldred{72.96}  \\ \bottomrule
% & RealOSR (Ours)      & \boldred{22.24} & \boldred{0.5948}  & \boldred{0.3131} & \boldred{0.1128}  & \boldred{51.63}   & 3.4158  & \boldblue{69.26} & 0.4219 & \boldblue{0.6065}  \\ \bottomrule
\end{tabular}
}
\label{tab:low_light}
\vspace{-2mm}
\end{table}

\textbf{Limitations.} The proposed RealOSR depends on the computationally intensive Stable Diffusion, which currently restricts its deployment on edge devices. Future work could explore lightweight alternatives to facilitate deployment on resource-constrained devices, thereby broadening RealOSR’s practical applicability across diverse environments and platforms.

\section{Conclusion}

In this paper, we introduced RealOSR, a diffusion-based approach for real-world omnidirectional image super-resolution (Real-ODISR), which provides more effective and efficient condition guidance for real-world degradations than previous methods. RealOSR integrates real-world degradation into the Latent Gradient Alignment Routing within a one-step denoising framework, empirically demonstrating enhanced fidelity and realness in super-resolution results. These advancements not only improve performance on constructed benchmarks but also establish a strong baseline model for future research in Real-ODISR. Further optimization of model efficiency could enable real-time applications in areas such as virtual reality and live broadcasting.

\vspace{-5pt}
\bibliography{ref}
\bibliographystyle{IEEEtran}

\end{document}